\documentclass[aps,prl,reprint,superscriptaddress]{revtex4-2}
\usepackage{graphicx}
\usepackage{color,amssymb,amsmath}
\usepackage[normalem]{ulem}
\usepackage{hyperref}
\usepackage{lipsum}

\newcommand{\Ic}{I_{\mathrm{C}}}
\newcommand{\Isw}{I_{\mathrm{SW}}}
\newcommand{\Icone}{I_{\mathrm{C,1}}}
\newcommand{\Ictwo}{I_{\mathrm{C,2}}}
\newcommand{\IcS}{I_{\mathrm{C,S}}}
\newcommand{\Imone}{I_{\mathrm{M,1}}}
\newcommand{\Imtwo}{I_{\mathrm{M,2}}}
\newcommand{\ImS}{I_{\mathrm{M,S}}}
\newcommand{\sone}{\sigma_{\mathrm{1}}}
\newcommand{\stwo}{\sigma_{\mathrm{2}}}
\newcommand{\sS}{\sigma_{\mathrm{S}}}
\newcommand{\Idc}{I_{\mathrm{DC}}}
\newcommand{\Vgone}{V_{\mathrm{G1}}}
\newcommand{\Vgtwo}{V_{\mathrm{G2}}}
\newcommand{\Vglob}{V_{\mathrm{Global}}}

\newcommand{\Imean}{I_{\mathrm{M}}}
\newcommand{\Ir}{I_{\mathrm{R}}}

\newcommand{\dU}{\Delta U}
\newcommand{\plasma}{\omega_{\mathrm{P}}}
\newcommand{\Bperp}{B_{\perp}}
\newcommand{\Tstar}{T^{*}}
\newcommand{\Ej}{E_{\mathrm{J}}}
\newcommand{\Ec}{E_{\mathrm{C}}}

\newcommand{\GQ}{\Gamma_{\mathrm{Q}}}
\newcommand{\GI}{\Gamma_{\mathrm{I}}}
\newcommand{\GR}{\Gamma_{\mathrm{R}}}
\newcommand{\GT}{\Gamma_{\mathrm{T}}}
\begin{document}
\title{Measurements of Phase Dynamics in Planar Josephson Junctions and SQUIDs}

\author{D. Z. Haxell}
\affiliation{IBM Research Europe - Zurich, S\"aumerstrasse 4, 8803 R\"uschlikon, Switzerland}

\author{E. Cheah}
\affiliation{Solid State Physics Laboratory, ETH Zurich, 8093 Zurich, Switzerland}

\author{F. Krizek}
\affiliation{IBM Research Europe - Zurich, S\"aumerstrasse 4, 8803 R\"uschlikon, Switzerland}
\affiliation{Solid State Physics Laboratory, ETH Zurich, 8093 Zurich, Switzerland}

\author{R. Schott}
\affiliation{Solid State Physics Laboratory, ETH Zurich, 8093 Zurich, Switzerland}

\author{M. F. Ritter}
\affiliation{IBM Research Europe - Zurich, S\"aumerstrasse 4, 8803 R\"uschlikon, Switzerland}

\author{M. Hinderling}
\affiliation{IBM Research Europe - Zurich, S\"aumerstrasse 4, 8803 R\"uschlikon, Switzerland}

\author{W. Belzig}
\affiliation{Fachbereich Physik, Universit\"at Konstanz, D-78457 Konstanz, Germany}

\author{C. Bruder}
\affiliation{Department of Physics, University of Basel, Klingelbergstrasse 82, CH-4056 Basel, Switzerland}

\author{W. Wegscheider}
\affiliation{Solid State Physics Laboratory, ETH Zurich, 8093 Zurich, Switzerland}

\author{H. Riel}
\affiliation{IBM Research Europe - Zurich, S\"aumerstrasse 4, 8803 R\"uschlikon, Switzerland}

\author{F. Nichele}
\email{fni@zurich.ibm.com}
\affiliation{IBM Research Europe - Zurich, S\"aumerstrasse 4, 8803 R\"uschlikon, Switzerland}

\date{\today}

\begin{abstract}
We experimentally investigate the stochastic phase dynamics of planar Josephson junctions (JJs) and superconducting quantum interference devices (SQUIDs) defined in epitaxial InAs/Al heterostructures, and characterized by a large ratio of Josephson energy to charging energy. We observe a crossover from a regime of macroscopic quantum tunneling to one of phase diffusion as a function of temperature, where the transition temperature $T^{*}$ is gate-tunable. The switching probability distributions are shown to be consistent with a small shunt capacitance and moderate damping, resulting in a switching current which is a small fraction of the critical current. Phase locking between two JJs leads to a difference in switching current between that of a JJ measured in isolation and that of the same JJ measured in an asymmetric SQUID loop. In the case of the loop, $T^*$ is also tuned by a magnetic flux.
\end{abstract}

\maketitle

Two-dimensional superconductor/semiconductor hybrid systems are a promising platform for scalable quantum computation and for the study of novel physical phenomena. The possibility to produce transparent interfaces~\cite{Krogstrup2015,Chang2015,Shabani2016,Vigneau2019,Moehle2021,Perla2021,Kanne2021,Pendharkar2021,Aggarwal2021}, combined with flexible lithographic patterning, is paving the way to a new generation of voltage-tunable qubit architectures~\cite{Larsen2015,Casparis2018, Janvier2015, deLange2015,Vidal2020,Wang2019,Hays2021}, with planar Josephson junctions (JJs) and superconducting quantum interference devices (SQUIDs) as core elements. Furthermore, spin-orbit interaction and Zeeman fields enable a rich playground for fundamental physics~\cite{Hart2017,Mayer2020,Baumgartner2022}, including the realization of topological states of matter~\cite{Hell2017,Pientka2017,Fornieri2019,Ren2019,Dartiailh2021,Banerjee2022}. In this context, understanding the phase dynamics of hybrid JJs and SQUIDs, which ultimately determine their switching currents, is crucial.

Here we investigate the stochastic phase dynamics of hybrid JJs defined in an InAs/Al planar heterostructure~\cite{Shabani2016}. We show that macroscopic quantum tunneling (MQT) and phase diffusion (PD) are the most relevant phase escape regimes. The low-temperature mean switching current $\Imean$ is a small fraction of the critical current $\Ic$, although the Josephson energy $\Ej$ is significantly larger than the charging energy $\Ec$. In JJs with small $\Ic$, the suppression of $\Imean$ is strong enough that PD dominates at low temperature. Embedding a JJ in an asymmetric SQUID, an approach intensively pursued for realizing topological states~\cite{Fornieri2019,Ren2019,Dartiailh2021,Banerjee2022}, modifies the phase escape mechanism. Thus, $\Imean$ may significantly vary when a JJ is measured in isolation or in a SQUID (by a factor of approximately 2.5, in the present case). The dominant phase-escape mechanism is further tuned via temperature, gate voltages and fluxes threading the SQUID. Contrary to conventional metallic JJs, no indication of thermal phase activation is observed. Characteristic experimental features are reproduced with a Monte Carlo simulation of the phase dynamics. Our results indicate that phase dynamics significantly affect the switching current of hybrid devices, and guide towards the realization of novel quantum architectures.

\begin{figure}
	\includegraphics[width=\columnwidth]{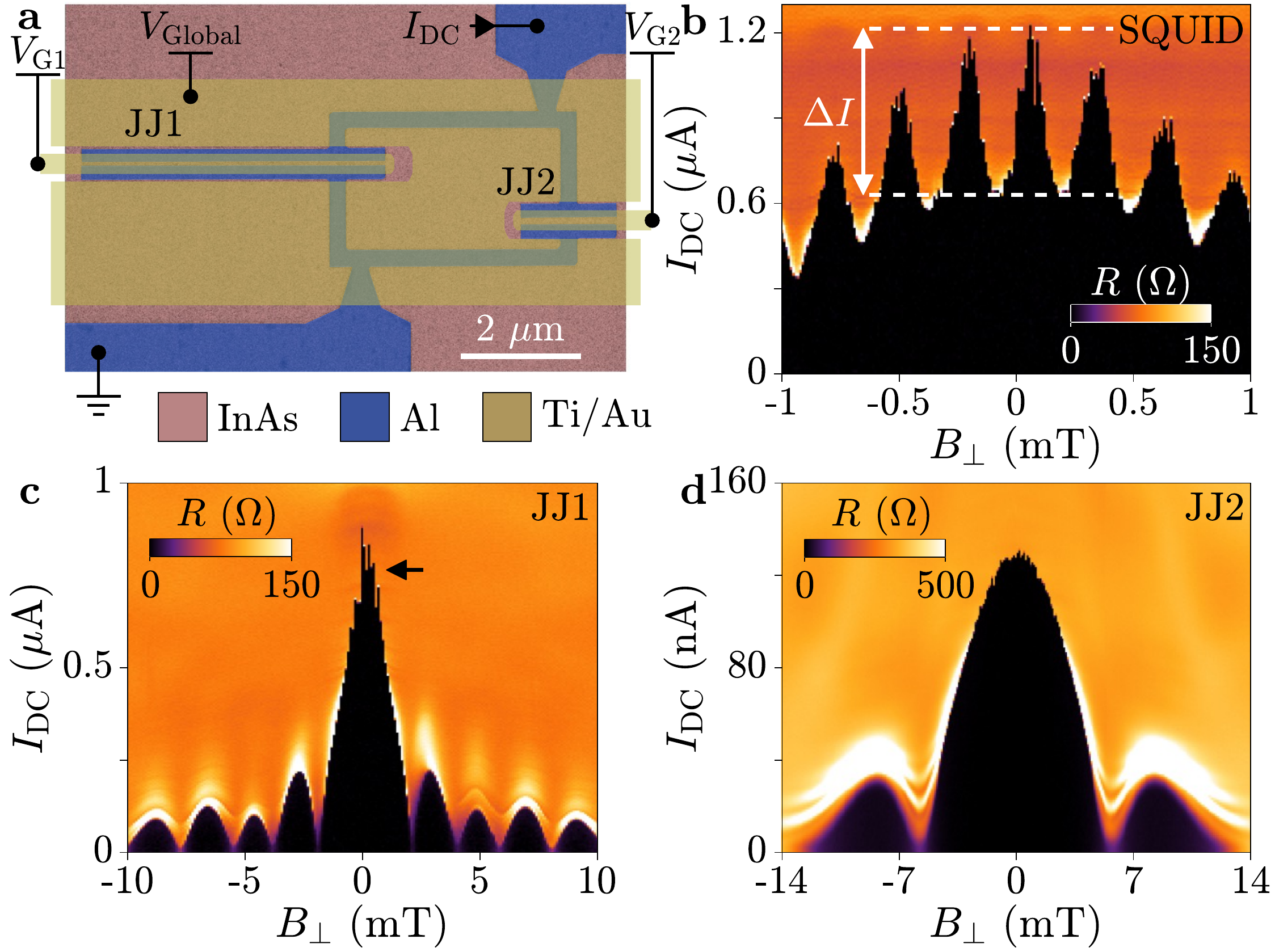}
	\caption{(a) False-colored electron micrograph of the device under study and measurement configuration. The InAs is highlighted in pink and the Al in blue. Gates are drawn on the image and highlighted in yellow. (b) Differential resistance $R$ as a function of $\Bperp$ and $\Idc$ obtained with $\Vgone=-180~\mathrm{mV}$ and $\Vgtwo=-140~\mathrm{mV}$. The amplitude of the switching current oscillations, $\Delta I$, is marked. (c) Differential resistance of JJ1 in isolation, with $\Vgone=-180~\mathrm{mV}$ and $\Vgtwo=-450~\mathrm{mV}$. Large fluctuations close to $\Bperp=0$ are marked with an arrow. (d) Differential resistance of JJ2 in isolation, with $\Vgone=-550~\mathrm{mV}$ and $\Vgtwo=-140~\mathrm{mV}$. The peak at $\Bperp=0$ is less than half $\Delta I/2$ in (b).}
	\label{fig1}
\end{figure}

\begin{figure*}
	\includegraphics[width=\textwidth]{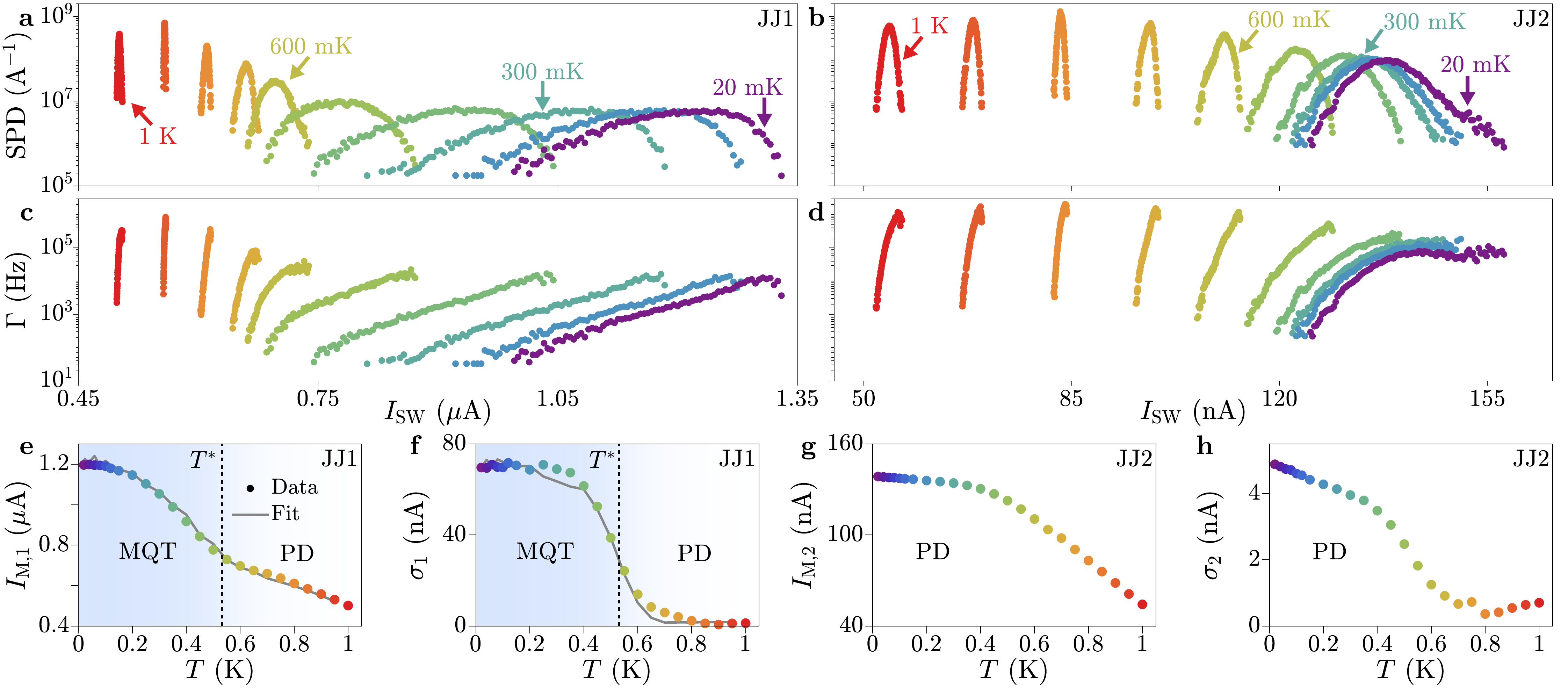}
	\caption{(a) Switching probability distributions (SPDs) for JJ1 for various temperatures. Colors are defined in (e) and are consistent throughout the text. (b) As in (a), but for JJ2. (c) Escape rate $\Gamma$ of JJ1, obtained from the data in (a) using Eq.~(\ref{eq:1}). (d) As in (c), but for JJ2. (e) Mean switching current $\Isw$ of SPDs for JJ1 as a function of temperature (circles) together with a fit to a Monte Carlo simulation (line). Transition temperature $\Tstar$ is indicated by a vertical line, dividing a regime of MQT (blue shading) and PD. (f) Standard deviation $\sone$ of SPDs for JJ1, as a function of temperature. (g,h) As in (e) and (f), for JJ2.}
	\label{fig2}
\end{figure*}

Figure~\ref{fig1}(a) shows a micrograph of the device under study. It consists of two gate-tunable planar JJs (JJ1 and JJ2) embedded in a SQUID loop, all defined in an InAs quantum well (pink) covered by a thin layer of in-situ-deposited Al (blue)~\cite{Shabani2016}. Devices were defined by wet etching of the Al, followed by deposition of a $15~\mathrm{nm}$ $\mathrm{HfO_x}$ layer and metallic gates (yellow). Gate voltages $\Vgone$ and $\Vgtwo$ allowed tuning of JJ1 and JJ2, respectively. The gate voltage $\Vglob$ was kept constant at $-600~\mathrm{mV}$ to prevent parallel conduction in the semiconductor. The design was optimized to reach a critical current in JJ1 ($\Icone$) that was much larger than the critical current in JJ2 ($\Ictwo$)~\cite{Fornieri2019,Nichele2020}. This was achieved by changing the lateral extent of the Al electrodes ($5~\mathrm{\mu m}$ in JJ1 vs. $1.6~\mathrm{\mu m}$ in JJ2) and their separation ($50~\mathrm{nm}$ in JJ1 vs. $100~\mathrm{nm}$ in JJ2). Electronic measurements were conducted in a dilution refrigerator with a mixing chamber base temperature below $20~\mathrm{mK}$. Results presented here were confirmed on a second SQUID device and on several individual JJs.

We first present switching currents obtained with low-frequency lock-in techniques, similar to previous work~\cite{Fornieri2019,Nichele2020,Mayer2020}. A source-drain current $\Idc$ was swept over timescales of seconds, while the SQUID differential resistance $R$ was recorded. Figure~\ref{fig1}(b) shows $R$ as a function of out-of-plane magnetic field $\Bperp$ with $\Vgone=-180~\mathrm{mV}$ and $\Vgtwo=-140~\mathrm{mV}$, where $\Icone$ and $\Ictwo$ were independently maximized. The SQUID switching current $I$ had a periodicity of $350~\mathrm{\mu T}$, corresponding to a flux $h/2e$ threading the loop. The amplitude of the SQUID oscillations, $\Delta I$, reveals the switching current of JJ2 as $I_{2}=\Delta I/2 = 350~\mathrm{nA}$, while the mean value gives the switching current of JJ1, $I_{1}=850~\mathrm{nA}$. Figure~\ref{fig1}(c) shows $R$ when JJ2 is closed and with JJ1 in the gate configuration of Fig.~\ref{fig1}(b). The Fraunhofer interference pattern emerges~\cite{Suominen2017}, with a maximum of $I_{1}$ matching the mean switching current of Fig.~\ref{fig1}(b). Furthermore, large switching current fluctuations were present at $\Bperp=0$ (black arrow). Figure~\ref{fig1}(d) shows similar measurements performed with $\Idc$ flowing in JJ2 only. Surprisingly, the maximum of $I_{2}$ is $120~\mathrm{nA}$; a significant difference with the $350~\mathrm{nA}$ deduced from Fig.~\ref{fig1}(b).

\begin{figure*}
	\includegraphics[width=\textwidth]{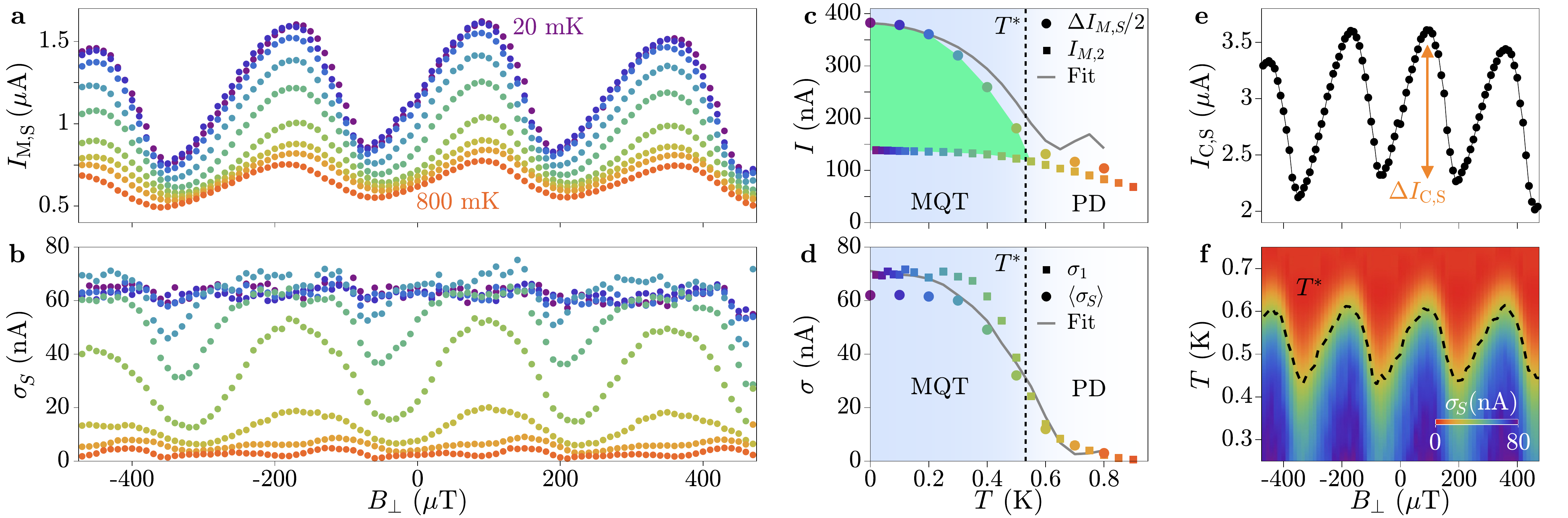}
	\caption{Mean $\ImS$ (a) and standard deviation $\sS$ (b) of the SPDs in the SQUID configuration as a function of $\Bperp$, for temperatures between $20$ and $800~\mathrm{mK}$. (c) Mean switching current of JJ2 as a function of $T$, derived from the SQUID oscillations (circles) and measured with JJ2 in isolation (squares). The solid line is $\Delta \ImS/2$ obtained from a Monte Carlo simulation fitted to the experimental results. (d) Standard deviation of the SPD in JJ1 measured in isolation (squares) together with the mean of $\sS$ from (b) (circles) as a function of temperature. The solid line is the result of the Monte Carlo simulation presented in (c). (e) SQUID critical current obtained by fitting the SPDs for $T=20~\mathrm{mK}$ to an MQT escape rate. (f) Color map of fitted standard deviation $\sS$, with transition temperature $T^{*}$ marked by a dashed line.}
	\label{fig3}
\end{figure*}

\begin{figure}
 	\includegraphics[width=\columnwidth]{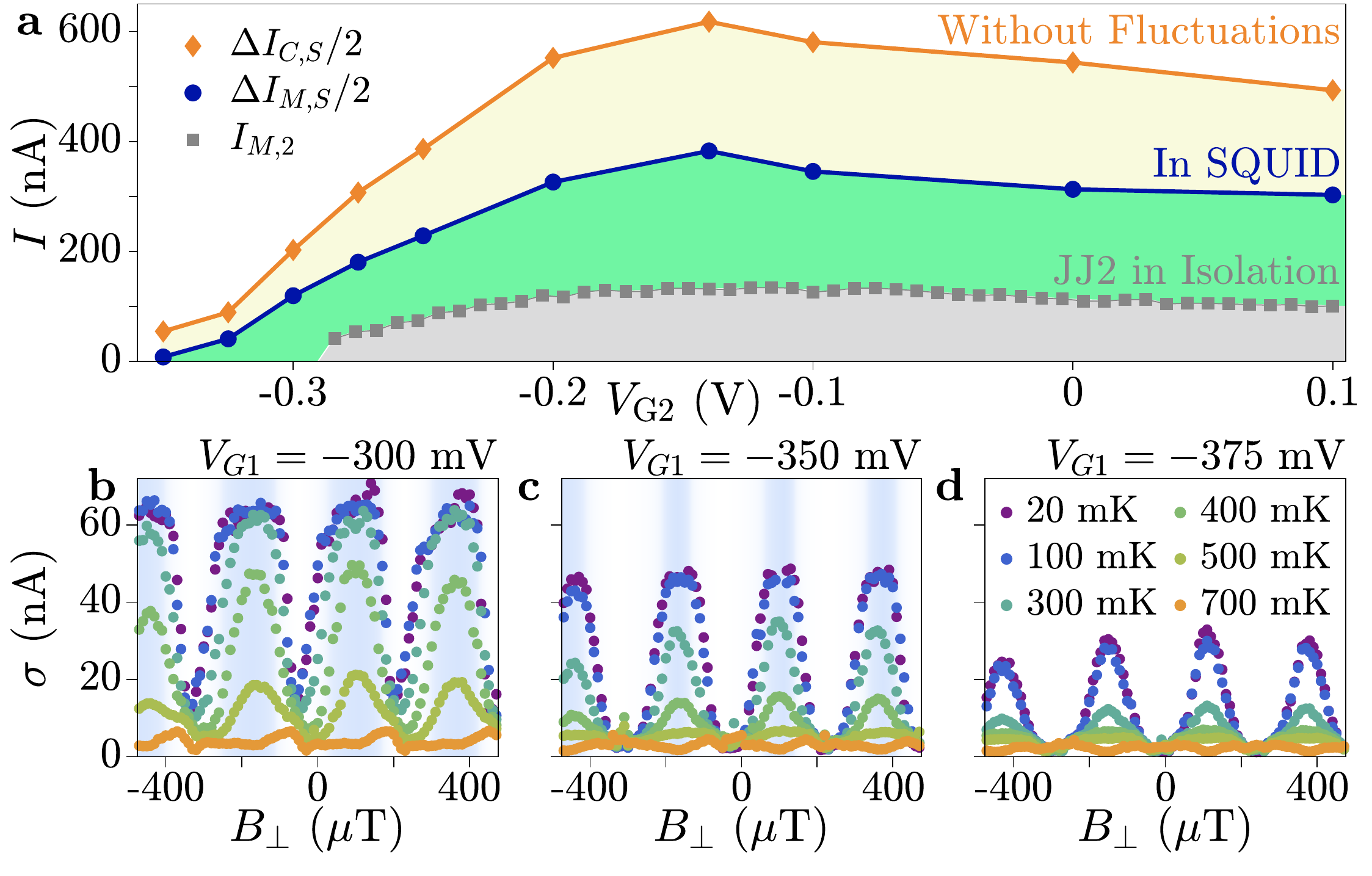}
 	\caption{(a) Switching currents of JJ2 as a function of $\Vgtwo$ when measured in isolation (squares) and in the SQUID configuration (circles), together with the critical current derived from Monte Carlo simulations (diamonds). (b-d) Standard deviation of SPDs measured in the SQUID configuration for three values of $\Vgone$. Blue shading highlights MQT regimes.}
 	\label{fig4}
 \end{figure}

Both the fluctuations in Fig.~\ref{fig1}(c) and the switching current enhancement in Fig.~\ref{fig1}(b) with respect to Fig.~\ref{fig1}(d) are manifestations of the phase dynamics in our devices. Therefore, we evaluate the phase escape mechanisms in JJ1 and JJ2 separately (Fig.~\ref{fig2}), and in the SQUID loop formed by their combination (Fig.~\ref{fig3}). Finally, we demonstrate gate and flux tunability of the escape dynamics (Fig.~\ref{fig4}). To capture the stochastic characteristics of phase escape, we modulate the input current with a sawtooth function using a ramp rate $\nu=240~\mathrm{\mu As^{-1}}$ and monitor the voltage across the SQUID with an oscilloscope. This technique allows us to record the switching current $\Isw$ for 10,000 switching events in approximately ten minutes, and produce the switching probability distribution (SPD), that is the probability for a switch to occur per unit of input current. Similar techniques were used for detailed studies of conventional~\cite{Fulton1974,Martinis1987} and hybrid JJs~\cite{Lee2011,Murphy2013,Kim2016,Kim2017}, metallic nanowires~\cite{Sahu2009,Li2011,Aref2012} and SQUIDs~\cite{LefevreSeguin1992, Li2002,Balestro2003,Sullivan2013,Butz2014}. 

Figures~\ref{fig2}(a) and (b) show the SPDs of JJ1 and JJ2, respectively, measured at various mixing chamber temperatures $T$. The corresponding escape rates $\Gamma$, computed as~\cite{Bezryadin2012}
\begin{equation}
	\Gamma(\Isw) = \mathrm{SPD}(\Isw) \nu \left[1- \int_{0}^{\Isw} \mathrm{SPD}(I)dI \right]^{-1},
	\label{eq:1}
\end{equation}
are shown in Fig.~\ref{fig2}(b) and (d), respectively.

Figure~\ref{fig2}(e) and (f) show the mean value of the SPDs in JJ1 ($\Imone$) and its standard deviation ($\sone$), respectively, both as a function of $T$. For $T<400~\mathrm{mK}$, $\sone$ is constant and large, and $\Gamma$ increases exponentially with $\Isw$, indicating that MQT dominates the phase dynamics. For higher $T$, $\sone$ decreases as $T$ increases, signaling the crossover to PD, where escape and retrapping events have similar probabilities to occur, so that many escape events are required to transition to the resistive state. The temperature $T^{*}\sim0.55~\mathrm{K}$ marks the crossover between a regime dominated by MQT and one dominated by PD. Regimes with $\sone$ increasing with $T$, which indicate thermal activation (TA), were not observed. The width of the low-temperature SPD, expressed as $\sigma/I_{1}=0.058$, is particularly large and results in pronounced switching current fluctuations, as seen in the measurements of Fig.~\ref{fig1}(c) (black arrow). Broad SPDs at low $T$, together with the absence of an intermediate TA regime, which is unusual in conventional JJs~\cite{Longobardi2012}, indicate a large critical current $\Icone$ and a small capacitance $C$ for JJ1. Finally, the relevance of PD, together with measuring a finite resistance at $\Idc=0$ for $T>1~\mathrm{K}$, which is well below the critical temperature $T_{\mathrm{C}}$ of the Al~\cite{Supplement}, indicates moderate damping.

The temperature dependence of $\Imone$ and $\sone$ is well captured by a Monte Carlo simulation of the phase dynamics [gray line in Figs.~\ref{fig2}(e,f)]~\cite{Fenton2008}, an approach previously adopted for the study of moderately damped JJs~\cite{Longobardi2011,Longobardi2011a}. In particular, the capacitance $C$ and the zero-temperature critical current $\Icone$ of JJ1 are first obtained by comparing the low-temperature data to a model of MQT. The quality factor $Q_0$ is subsequently determined by comparing the full temperature dependence to the Monte Carlo simulation. Details on our procedure are discussed in the Supplemental Material~\cite{Supplement}. While this model was developed for tunneling JJs with sinusoidal current-phase relation and large $Q_0$, in the absence of a more complete theory, we tentatively apply it to our devices and consider the results to be of qualitative nature.
The best fit was obtained with $C=1~\mathrm{fF}$, $\Icone=3~\mathrm{\mu A}$ and $Q_0=7$. As expected, JJ1 is moderately damped and has a small intrinsic capacitance, leading to a large plasma frequency. The estimated $\Icone$ is 2.5 times higher than $\Imone$, indicating that moderate input currents already result in a high switching probability. The ratio between $\Icone$ and $\Imone$ decreases towards one for $T>T^{*}$, as signaled by the kink in $\Imone$ at $T=T^{*}$. The result $C=1~\mathrm{fF}$ is consistent with the geometrical capacitance between the Al electrodes~\cite{Supplement}. With these parameters, we estimate $\Ej/\Ec=73$ at $T=20~\mathrm{mK}$~\cite{Supplement}. This situation is very different from conventional metallic Josephson junction, where strong suppression of $\Isw$ from $\Ic$ requires $\Ej/\Ec\leq1$~\cite{Martinis1989,Iansiti1987,Iansiti1989}. Due to the small $C$ and large $\Ic$, we estimate that the transition from MQT to TA would occur for $T>T_{\mathrm{C}}$ so that, in the entire PD regime, phase escape takes place via MQT~\cite{Supplement}.

Similar to JJ1, MQT is the dominant phase escape mechanism in JJ2. However, large dissipation results in a significant retrapping probability and places JJ2 in the PD regime down to base temperature. This is evident from $\Imtwo$ and $\stwo$ shown in Figs.~\ref{fig2}(g) and (h), respectively, where $\stwo$ does not saturate for $T\rightarrow 0$, and from the deviation of $\Gamma$ from an exponential in Fig.~\ref{fig2}(d). The small $\Ictwo$ likely sets $Q_0\sim1$, which is outside the range of validity of our Monte Carlo simulations.

We now present the phase dynamics when both JJs are activated. Figures~\ref{fig3}(a) and (b) show the mean, $\ImS$, and standard deviation, $\sS$, of each SPD obtained in the gate configuration of Fig.~\ref{fig1}(b) as a function of $\Bperp$ and $T$~\footnote{Each SPD was obtained by recording 5,000 switching events.}. In Fig.~\ref{fig3}(a), SQUID oscillations are clearly captured by $\ImS$. In Fig.~\ref{fig3}(b), the curves at low $T$ have a large $\sS$, independent of $\Bperp$. As $T$ increases further, $\sS$ is modulated by $\Bperp$ and ultimately becomes small and independent of $\Bperp$. In Fig.~\ref{fig3}(c) we compare $\Imtwo$ [squares, as in Fig.~\ref{fig2}(g)] to the half-amplitude of the oscillations in $\ImS$ (circles). In the absence of macroscopic quantum tunneling, the two quantities would coincide. Instead, we find a significant discrepancy, highlighted by green shading, which is large at low $T$ and vanishes above $T^{*}$ of JJ1. By tuning $T^{*}$ via $\Vgone$, we confirm that the enhancement of $\Delta \ImS/2$ with respect to $\Imtwo$ was always correlated to $T^{*}$ in JJ1~\cite{Supplement}. The mean value of $\ImS$ matched $\Imone$~\cite{Supplement} and the mean of $\sS$, $\langle\sS\rangle$, was similar to $\sone$ [Fig.~\ref{fig3}(d)].

The results presented in Fig.~\ref{fig3} are intuitively understood by considering phase-locking by the loop inductance. For JJ2 alone, phase escape is more likely at moderate currents compared to JJ1. Coupling JJ2 to JJ1 effectively realizes a new JJ with higher Josephson energy and similar phase dynamics to JJ1, so that the dominant switching mechanism is MQT and, consequently, the suppression of $\Imean$ is reduced. However, protection of $\Imtwo$ is maintained while JJ1 stays in the MQT regime ($T<T^{*}$), where phase uncertainty is less than in the PD regime.
Consistent with this interpretation, phase dynamics in the asymmetric SQUID configuration are well described by a Monte Carlo simulation of a fictitious JJ with a field-dependent critical current $\IcS(\Bperp)$, and with $C$ and $Q_{0}$ as derived for JJ1. The sole fit parameter was $\IcS$ for $T=20~\mathrm{mK}$, which is shown in Fig.~\ref{fig3}(e) as a function of $\Bperp$ (circles). The curve is consistent with the presence of highly-transmissive Andreev bound states (ABSs), resulting in a forward-skewed current-phase relation~\cite{Beenakker1991,Nichele2020}. Also for the SQUID, critical current $\IcS$ and mean switching current $\ImS$ [Fig.~\ref{fig3}(a)] differ by a factor of approximately 2.5. After obtaining $\IcS(\Bperp)$ for $T=20~\mathrm{mK}$, the entire dataset of Figs.~\ref{fig3}(a) and (b) was simulated without free parameters. We show the simulated half-amplitude $\Delta \ImS/2$ and the mean of $\sS$ as gray lines in Figs.~\ref{fig3}(c) and (d) respectively. Despite the simplicity of our model, experimental results are reproduced to a large extent. Figure~\ref{fig3}(f) shows a colormap of the simulated standard deviation, $\sS(\Bperp,T)$, with $T^{*}$ indicated by a dashed line and marking the crossover between MQT and PD. The phase dynamics are completely described by MQT and PD for low and high $T$, respectively. For intermediate $T$, the phase escape mechanism periodically varies between MQT and PD as a function of $\Bperp$.

In the following, we discuss how phase escape dynamics vary as $\Icone$ and $\Ictwo$ are tuned via gate voltages. Figure~\ref{fig4}(a) summarizes results for JJ2 as $\Vgtwo$ was varied. When JJ2 was measured in isolation, switching currents $\Imtwo$ were small and PD was the dominant regime throughout the accessible range of $\Vgtwo$. We highlight this condition with gray shading. When the SQUID was formed, the switching current of JJ2 deduced from the SQUID oscillations ($\Delta \IcS/2$) was significantly higher than when JJ2 was measured in isolation. We highlight this situation with green shading. For $\Vgtwo<300~\mathrm{mV}$ JJ2 was resistive, if measured in isolation, presumably due to $\Ej/\Ec\approx1$~\cite{Martinis1989,Iansiti1987,Iansiti1989}, but SQUID oscillations were still observed. Finally, the $\Ictwo$ obtained by fitting the SPDs in the SQUID with the Monte Carlo simulation [as in Fig.~\ref{fig3}(e)] is highlighted in yellow.

Decreasing $\Icone$ via $\Vgone$ made the SQUID more symmetric and shifted JJ1 towards a regime of PD. Figures~\ref{fig4}(b-d) show $\sS$ for decreasing values of $\Vgone$. For $\Vgone=-300~\mathrm{mV}$ and $-350~\mathrm{mV}$ (Figs.~\ref{fig4}(c) and (d), respectively) escape dynamics varied between MQT (blue shading) and PD already at base temperature, with $\sS$ oscillating between $10$ and $60~\mathrm{nA}$ within one SQUID oscillation. For $\Vgone=-375~\mathrm{mV}$ [Fig.~\ref{fig4}(d)] PD dominated at low $T$, although modulations in $\sS$ persisted. Modeling the curves in Figs.~\ref{fig4}(b-d) would require a quantum treatment of the phase escape from a 2D potential, which goes beyond the scope of this work.

In conclusion, the switching current of planar JJs with highly transmissive ABSs is strongly affected by phase dynamics, even for $\Ej/\Ec\gg1$. As a result of moderate dissipation ($Q_{0}<10$) and large plasma frequency, $\Imean$ can differ largely from $\Ic$, depending on the details of the JJs and of their electrostatic environment. Phase dynamics can be modified by embeddeding JJs in asymmetric SQUID geometries, resulting in significant changes of $\Imean$. Furthermore, the dominant phase-escape mechanism in a SQUID can be tuned between MQT and PD via a magnetic flux, affecting the SQUID switching current and its standard deviation. This intricate physics is relevant for realizing gate-tunable quantum devices and investigating topological phenomena, where hybrid JJs with phase control are widespread.

\begin{acknowledgments}
We are grateful to C.~M\"{u}ller and W.~Riess for helpful discussions. We thank the Cleanroom Operations Team of the Binnig and Rohrer Nanotechnology Center (BRNC) for their help and support. F.~N. acknowledges support from the European Research Council (grant number 804273) and the Swiss National Science Foundation (grant number 200021\_201082). W.~B. acknowledges support from the European Union's Horizon 2020 FET Open programme (grant number 964398) and from  the Deutsche Forschungsgemeinschaft (DFG; German Research Foundation) via the SFB 1432 (ID 425217212).
\end{acknowledgments}

\bibliography{Bibliography}

\begin{thebibliography}{58}%
\makeatletter
\providecommand \@ifxundefined [1]{%
 \@ifx{#1\undefined}
}%
\providecommand \@ifnum [1]{%
 \ifnum #1\expandafter \@firstoftwo
 \else \expandafter \@secondoftwo
 \fi
}%
\providecommand \@ifx [1]{%
 \ifx #1\expandafter \@firstoftwo
 \else \expandafter \@secondoftwo
 \fi
}%
\providecommand \natexlab [1]{#1}%
\providecommand \enquote  [1]{``#1''}%
\providecommand \bibnamefont  [1]{#1}%
\providecommand \bibfnamefont [1]{#1}%
\providecommand \citenamefont [1]{#1}%
\providecommand \href@noop [0]{\@secondoftwo}%
\providecommand \href [0]{\begingroup \@sanitize@url \@href}%
\providecommand \@href[1]{\@@startlink{#1}\@@href}%
\providecommand \@@href[1]{\endgroup#1\@@endlink}%
\providecommand \@sanitize@url [0]{\catcode `\\12\catcode `\$12\catcode
  `\&12\catcode `\#12\catcode `\^12\catcode `\_12\catcode `\%12\relax}%
\providecommand \@@startlink[1]{}%
\providecommand \@@endlink[0]{}%
\providecommand \url  [0]{\begingroup\@sanitize@url \@url }%
\providecommand \@url [1]{\endgroup\@href {#1}{\urlprefix }}%
\providecommand \urlprefix  [0]{URL }%
\providecommand \Eprint [0]{\href }%
\providecommand \doibase [0]{https://doi.org/}%
\providecommand \selectlanguage [0]{\@gobble}%
\providecommand \bibinfo  [0]{\@secondoftwo}%
\providecommand \bibfield  [0]{\@secondoftwo}%
\providecommand \translation [1]{[#1]}%
\providecommand \BibitemOpen [0]{}%
\providecommand \bibitemStop [0]{}%
\providecommand \bibitemNoStop [0]{.\EOS\space}%
\providecommand \EOS [0]{\spacefactor3000\relax}%
\providecommand \BibitemShut  [1]{\csname bibitem#1\endcsname}%
\let\auto@bib@innerbib\@empty
\bibitem [{\citenamefont {Krogstrup}\ \emph {et~al.}(2015)\citenamefont
  {Krogstrup}, \citenamefont {Ziino}, \citenamefont {Chang}, \citenamefont
  {Albrecht}, \citenamefont {Madsen}, \citenamefont {Johnson}, \citenamefont
  {Nygård}, \citenamefont {Marcus},\ and\ \citenamefont
  {Jespersen}}]{Krogstrup2015}%
  \BibitemOpen
  \bibfield  {author} {\bibinfo {author} {\bibfnamefont {P.}~\bibnamefont
  {Krogstrup}}, \bibinfo {author} {\bibfnamefont {N.~L.~B.}\ \bibnamefont
  {Ziino}}, \bibinfo {author} {\bibfnamefont {W.}~\bibnamefont {Chang}},
  \bibinfo {author} {\bibfnamefont {S.~M.}\ \bibnamefont {Albrecht}}, \bibinfo
  {author} {\bibfnamefont {M.~H.}\ \bibnamefont {Madsen}}, \bibinfo {author}
  {\bibfnamefont {E.}~\bibnamefont {Johnson}}, \bibinfo {author} {\bibfnamefont
  {J.}~\bibnamefont {Nygård}}, \bibinfo {author} {\bibfnamefont
  {C.}~\bibnamefont {Marcus}},\ and\ \bibinfo {author} {\bibfnamefont {T.~S.}\
  \bibnamefont {Jespersen}},\ }\href {https://doi.org/10.1038/nmat4176}
  {\bibfield  {journal} {\bibinfo  {journal} {Nature Materials}\ }\textbf
  {\bibinfo {volume} {14}},\ \bibinfo {pages} {400} (\bibinfo {year}
  {2015})}\BibitemShut {NoStop}%
\bibitem [{\citenamefont {Chang}\ \emph {et~al.}(2015)\citenamefont {Chang},
  \citenamefont {Albrecht}, \citenamefont {Jespersen}, \citenamefont
  {Kuemmeth}, \citenamefont {Krogstrup}, \citenamefont {Nygård},\ and\
  \citenamefont {Marcus}}]{Chang2015}%
  \BibitemOpen
  \bibfield  {author} {\bibinfo {author} {\bibfnamefont {W.}~\bibnamefont
  {Chang}}, \bibinfo {author} {\bibfnamefont {S.~M.}\ \bibnamefont {Albrecht}},
  \bibinfo {author} {\bibfnamefont {T.~S.}\ \bibnamefont {Jespersen}}, \bibinfo
  {author} {\bibfnamefont {F.}~\bibnamefont {Kuemmeth}}, \bibinfo {author}
  {\bibfnamefont {P.}~\bibnamefont {Krogstrup}}, \bibinfo {author}
  {\bibfnamefont {J.}~\bibnamefont {Nygård}},\ and\ \bibinfo {author}
  {\bibfnamefont {C.~M.}\ \bibnamefont {Marcus}},\ }\href
  {https://doi.org/10.1038/nnano.2014.306} {\bibfield  {journal} {\bibinfo
  {journal} {Nature Nanotechnology}\ }\textbf {\bibinfo {volume} {10}},\
  \bibinfo {pages} {232} (\bibinfo {year} {2015})}\BibitemShut {NoStop}%
\bibitem [{\citenamefont {Shabani}\ \emph {et~al.}(2016)\citenamefont
  {Shabani}, \citenamefont {Kjaergaard}, \citenamefont {Suominen},
  \citenamefont {Kim}, \citenamefont {Nichele}, \citenamefont {Pakrouski},
  \citenamefont {Stankevic}, \citenamefont {Lutchyn}, \citenamefont
  {Krogstrup}, \citenamefont {Feidenhans'l}, \citenamefont {Kraemer},
  \citenamefont {Nayak}, \citenamefont {Troyer}, \citenamefont {Marcus},\ and\
  \citenamefont {Palmstr\o{}m}}]{Shabani2016}%
  \BibitemOpen
  \bibfield  {author} {\bibinfo {author} {\bibfnamefont {J.}~\bibnamefont
  {Shabani}}, \bibinfo {author} {\bibfnamefont {M.}~\bibnamefont {Kjaergaard}},
  \bibinfo {author} {\bibfnamefont {H.~J.}\ \bibnamefont {Suominen}}, \bibinfo
  {author} {\bibfnamefont {Y.}~\bibnamefont {Kim}}, \bibinfo {author}
  {\bibfnamefont {F.}~\bibnamefont {Nichele}}, \bibinfo {author} {\bibfnamefont
  {K.}~\bibnamefont {Pakrouski}}, \bibinfo {author} {\bibfnamefont
  {T.}~\bibnamefont {Stankevic}}, \bibinfo {author} {\bibfnamefont {R.~M.}\
  \bibnamefont {Lutchyn}}, \bibinfo {author} {\bibfnamefont {P.}~\bibnamefont
  {Krogstrup}}, \bibinfo {author} {\bibfnamefont {R.}~\bibnamefont
  {Feidenhans'l}}, \bibinfo {author} {\bibfnamefont {S.}~\bibnamefont
  {Kraemer}}, \bibinfo {author} {\bibfnamefont {C.}~\bibnamefont {Nayak}},
  \bibinfo {author} {\bibfnamefont {M.}~\bibnamefont {Troyer}}, \bibinfo
  {author} {\bibfnamefont {C.~M.}\ \bibnamefont {Marcus}},\ and\ \bibinfo
  {author} {\bibfnamefont {C.~J.}\ \bibnamefont {Palmstr\o{}m}},\ }\href
  {https://doi.org/10.1103/PhysRevB.93.155402} {\bibfield  {journal} {\bibinfo
  {journal} {Phys. Rev. B}\ }\textbf {\bibinfo {volume} {93}},\ \bibinfo
  {pages} {155402} (\bibinfo {year} {2016})}\BibitemShut {NoStop}%
\bibitem [{\citenamefont {Vigneau}\ \emph {et~al.}(2019)\citenamefont
  {Vigneau}, \citenamefont {Mizokuchi}, \citenamefont {Zanuz}, \citenamefont
  {Huang}, \citenamefont {Tan}, \citenamefont {Maurand}, \citenamefont
  {Frolov}, \citenamefont {Sammak}, \citenamefont {Scappucci}, \citenamefont
  {Lefloch},\ and\ \citenamefont {De~Franceschi}}]{Vigneau2019}%
  \BibitemOpen
  \bibfield  {author} {\bibinfo {author} {\bibfnamefont {F.}~\bibnamefont
  {Vigneau}}, \bibinfo {author} {\bibfnamefont {R.}~\bibnamefont {Mizokuchi}},
  \bibinfo {author} {\bibfnamefont {D.~C.}\ \bibnamefont {Zanuz}}, \bibinfo
  {author} {\bibfnamefont {X.}~\bibnamefont {Huang}}, \bibinfo {author}
  {\bibfnamefont {S.}~\bibnamefont {Tan}}, \bibinfo {author} {\bibfnamefont
  {R.}~\bibnamefont {Maurand}}, \bibinfo {author} {\bibfnamefont
  {S.}~\bibnamefont {Frolov}}, \bibinfo {author} {\bibfnamefont
  {A.}~\bibnamefont {Sammak}}, \bibinfo {author} {\bibfnamefont
  {G.}~\bibnamefont {Scappucci}}, \bibinfo {author} {\bibfnamefont
  {F.}~\bibnamefont {Lefloch}},\ and\ \bibinfo {author} {\bibfnamefont
  {S.}~\bibnamefont {De~Franceschi}},\ }\href
  {https://doi.org/10.1021/acs.nanolett.8b04275} {\bibfield  {journal}
  {\bibinfo  {journal} {Nano Lett.}\ }\textbf {\bibinfo {volume} {19}},\
  \bibinfo {pages} {1023} (\bibinfo {year} {2019})}\BibitemShut {NoStop}%
\bibitem [{\citenamefont {Moehle}\ \emph {et~al.}(2021)\citenamefont {Moehle},
  \citenamefont {Ke}, \citenamefont {Wang}, \citenamefont {Thomas},
  \citenamefont {Xiao}, \citenamefont {Karwal}, \citenamefont {Lodari},
  \citenamefont {van~de Kerkhof}, \citenamefont {Termaat}, \citenamefont
  {Gardner}, \citenamefont {Scappucci}, \citenamefont {Manfra},\ and\
  \citenamefont {Goswami}}]{Moehle2021}%
  \BibitemOpen
  \bibfield  {author} {\bibinfo {author} {\bibfnamefont {C.~M.}\ \bibnamefont
  {Moehle}}, \bibinfo {author} {\bibfnamefont {C.~T.}\ \bibnamefont {Ke}},
  \bibinfo {author} {\bibfnamefont {Q.}~\bibnamefont {Wang}}, \bibinfo {author}
  {\bibfnamefont {C.}~\bibnamefont {Thomas}}, \bibinfo {author} {\bibfnamefont
  {D.}~\bibnamefont {Xiao}}, \bibinfo {author} {\bibfnamefont {S.}~\bibnamefont
  {Karwal}}, \bibinfo {author} {\bibfnamefont {M.}~\bibnamefont {Lodari}},
  \bibinfo {author} {\bibfnamefont {V.}~\bibnamefont {van~de Kerkhof}},
  \bibinfo {author} {\bibfnamefont {R.}~\bibnamefont {Termaat}}, \bibinfo
  {author} {\bibfnamefont {G.~C.}\ \bibnamefont {Gardner}}, \bibinfo {author}
  {\bibfnamefont {G.}~\bibnamefont {Scappucci}}, \bibinfo {author}
  {\bibfnamefont {M.~J.}\ \bibnamefont {Manfra}},\ and\ \bibinfo {author}
  {\bibfnamefont {S.}~\bibnamefont {Goswami}},\ }\href
  {https://doi.org/10.1021/acs.nanolett.1c03520} {\bibfield  {journal}
  {\bibinfo  {journal} {Nano Lett.}\ }\textbf {\bibinfo {volume} {21}},\
  \bibinfo {pages} {9990} (\bibinfo {year} {2021})}\BibitemShut {NoStop}%
\bibitem [{\citenamefont {Perla}\ \emph {et~al.}(2021)\citenamefont {Perla},
  \citenamefont {Fonseka}, \citenamefont {Zellekens}, \citenamefont {Deacon},
  \citenamefont {Han}, \citenamefont {Kölzer}, \citenamefont {Mörstedt},
  \citenamefont {Bennemann}, \citenamefont {Espiari}, \citenamefont
  {Ishibashi}, \citenamefont {Grützmacher}, \citenamefont {Sanchez},
  \citenamefont {Lepsa},\ and\ \citenamefont {Schäpers}}]{Perla2021}%
  \BibitemOpen
  \bibfield  {author} {\bibinfo {author} {\bibfnamefont {P.}~\bibnamefont
  {Perla}}, \bibinfo {author} {\bibfnamefont {H.~A.}\ \bibnamefont {Fonseka}},
  \bibinfo {author} {\bibfnamefont {P.}~\bibnamefont {Zellekens}}, \bibinfo
  {author} {\bibfnamefont {R.}~\bibnamefont {Deacon}}, \bibinfo {author}
  {\bibfnamefont {Y.}~\bibnamefont {Han}}, \bibinfo {author} {\bibfnamefont
  {J.}~\bibnamefont {Kölzer}}, \bibinfo {author} {\bibfnamefont
  {T.}~\bibnamefont {Mörstedt}}, \bibinfo {author} {\bibfnamefont
  {B.}~\bibnamefont {Bennemann}}, \bibinfo {author} {\bibfnamefont
  {A.}~\bibnamefont {Espiari}}, \bibinfo {author} {\bibfnamefont
  {K.}~\bibnamefont {Ishibashi}}, \bibinfo {author} {\bibfnamefont
  {D.}~\bibnamefont {Grützmacher}}, \bibinfo {author} {\bibfnamefont {A.~M.}\
  \bibnamefont {Sanchez}}, \bibinfo {author} {\bibfnamefont {M.~I.}\
  \bibnamefont {Lepsa}},\ and\ \bibinfo {author} {\bibfnamefont
  {T.}~\bibnamefont {Schäpers}},\ }\href {https://doi.org/10.1039/D0NA00999G}
  {\bibfield  {journal} {\bibinfo  {journal} {Nanoscale Adv.}\ }\textbf
  {\bibinfo {volume} {3}},\ \bibinfo {pages} {1413} (\bibinfo {year}
  {2021})}\BibitemShut {NoStop}%
\bibitem [{\citenamefont {Kanne}\ \emph {et~al.}(2021)\citenamefont {Kanne},
  \citenamefont {Marnauza}, \citenamefont {Olsteins}, \citenamefont {Carrad},
  \citenamefont {Sestoft}, \citenamefont {de~Bruijckere}, \citenamefont {Zeng},
  \citenamefont {Johnson}, \citenamefont {Olsson}, \citenamefont
  {Grove-Rasmussen},\ and\ \citenamefont {Nygård}}]{Kanne2021}%
  \BibitemOpen
  \bibfield  {author} {\bibinfo {author} {\bibfnamefont {T.}~\bibnamefont
  {Kanne}}, \bibinfo {author} {\bibfnamefont {M.}~\bibnamefont {Marnauza}},
  \bibinfo {author} {\bibfnamefont {D.}~\bibnamefont {Olsteins}}, \bibinfo
  {author} {\bibfnamefont {D.~J.}\ \bibnamefont {Carrad}}, \bibinfo {author}
  {\bibfnamefont {J.~E.}\ \bibnamefont {Sestoft}}, \bibinfo {author}
  {\bibfnamefont {J.}~\bibnamefont {de~Bruijckere}}, \bibinfo {author}
  {\bibfnamefont {L.}~\bibnamefont {Zeng}}, \bibinfo {author} {\bibfnamefont
  {E.}~\bibnamefont {Johnson}}, \bibinfo {author} {\bibfnamefont
  {E.}~\bibnamefont {Olsson}}, \bibinfo {author} {\bibfnamefont
  {K.}~\bibnamefont {Grove-Rasmussen}},\ and\ \bibinfo {author} {\bibfnamefont
  {J.}~\bibnamefont {Nygård}},\ }\href
  {https://doi.org/10.1038/s41565-021-00900-9} {\bibfield  {journal} {\bibinfo
  {journal} {Nature Nanotechnology}\ }\textbf {\bibinfo {volume} {16}},\
  \bibinfo {pages} {776} (\bibinfo {year} {2021})}\BibitemShut {NoStop}%
\bibitem [{\citenamefont {Pendharkar}\ \emph {et~al.}(2021)\citenamefont
  {Pendharkar}, \citenamefont {Zhang}, \citenamefont {Wu}, \citenamefont
  {Zarassi}, \citenamefont {Zhang}, \citenamefont {Dempsey}, \citenamefont
  {Lee}, \citenamefont {Harrington}, \citenamefont {Badawy}, \citenamefont
  {Gazibegovic}, \citenamefont {{Op het Veld}}, \citenamefont {Rossi},
  \citenamefont {Jung}, \citenamefont {Chen}, \citenamefont {Verheijen},
  \citenamefont {Hocevar}, \citenamefont {Bakkers}, \citenamefont
  {Palmstrøm},\ and\ \citenamefont {Frolov}}]{Pendharkar2021}%
  \BibitemOpen
  \bibfield  {author} {\bibinfo {author} {\bibfnamefont {M.}~\bibnamefont
  {Pendharkar}}, \bibinfo {author} {\bibfnamefont {B.}~\bibnamefont {Zhang}},
  \bibinfo {author} {\bibfnamefont {H.}~\bibnamefont {Wu}}, \bibinfo {author}
  {\bibfnamefont {A.}~\bibnamefont {Zarassi}}, \bibinfo {author} {\bibfnamefont
  {P.}~\bibnamefont {Zhang}}, \bibinfo {author} {\bibfnamefont {C.~P.}\
  \bibnamefont {Dempsey}}, \bibinfo {author} {\bibfnamefont {J.~S.}\
  \bibnamefont {Lee}}, \bibinfo {author} {\bibfnamefont {S.~D.}\ \bibnamefont
  {Harrington}}, \bibinfo {author} {\bibfnamefont {G.}~\bibnamefont {Badawy}},
  \bibinfo {author} {\bibfnamefont {S.}~\bibnamefont {Gazibegovic}}, \bibinfo
  {author} {\bibfnamefont {R.~L.~M.}\ \bibnamefont {{Op het Veld}}}, \bibinfo
  {author} {\bibfnamefont {M.}~\bibnamefont {Rossi}}, \bibinfo {author}
  {\bibfnamefont {J.}~\bibnamefont {Jung}}, \bibinfo {author} {\bibfnamefont
  {A.-H.}\ \bibnamefont {Chen}}, \bibinfo {author} {\bibfnamefont {M.~A.}\
  \bibnamefont {Verheijen}}, \bibinfo {author} {\bibfnamefont {M.}~\bibnamefont
  {Hocevar}}, \bibinfo {author} {\bibfnamefont {E.~P. A.~M.}\ \bibnamefont
  {Bakkers}}, \bibinfo {author} {\bibfnamefont {C.~J.}\ \bibnamefont
  {Palmstrøm}},\ and\ \bibinfo {author} {\bibfnamefont {S.~M.}\ \bibnamefont
  {Frolov}},\ }\href {https://doi.org/10.1126/science.aba5211} {\bibfield
  {journal} {\bibinfo  {journal} {Science}\ }\textbf {\bibinfo {volume}
  {372}},\ \bibinfo {pages} {508} (\bibinfo {year} {2021})}\BibitemShut
  {NoStop}%
\bibitem [{\citenamefont {Aggarwal}\ \emph {et~al.}(2021)\citenamefont
  {Aggarwal}, \citenamefont {Hofmann}, \citenamefont {Jirovec}, \citenamefont
  {Prieto}, \citenamefont {Sammak}, \citenamefont {Botifoll}, \citenamefont
  {Mart\'{\i}-S\'anchez}, \citenamefont {Veldhorst}, \citenamefont {Arbiol},
  \citenamefont {Scappucci}, \citenamefont {Danon},\ and\ \citenamefont
  {Katsaros}}]{Aggarwal2021}%
  \BibitemOpen
  \bibfield  {author} {\bibinfo {author} {\bibfnamefont {K.}~\bibnamefont
  {Aggarwal}}, \bibinfo {author} {\bibfnamefont {A.}~\bibnamefont {Hofmann}},
  \bibinfo {author} {\bibfnamefont {D.}~\bibnamefont {Jirovec}}, \bibinfo
  {author} {\bibfnamefont {I.}~\bibnamefont {Prieto}}, \bibinfo {author}
  {\bibfnamefont {A.}~\bibnamefont {Sammak}}, \bibinfo {author} {\bibfnamefont
  {M.}~\bibnamefont {Botifoll}}, \bibinfo {author} {\bibfnamefont
  {S.}~\bibnamefont {Mart\'{\i}-S\'anchez}}, \bibinfo {author} {\bibfnamefont
  {M.}~\bibnamefont {Veldhorst}}, \bibinfo {author} {\bibfnamefont
  {J.}~\bibnamefont {Arbiol}}, \bibinfo {author} {\bibfnamefont
  {G.}~\bibnamefont {Scappucci}}, \bibinfo {author} {\bibfnamefont
  {J.}~\bibnamefont {Danon}},\ and\ \bibinfo {author} {\bibfnamefont
  {G.}~\bibnamefont {Katsaros}},\ }\href
  {https://doi.org/10.1103/PhysRevResearch.3.L022005} {\bibfield  {journal}
  {\bibinfo  {journal} {Phys. Rev. Research}\ }\textbf {\bibinfo {volume}
  {3}},\ \bibinfo {pages} {L022005} (\bibinfo {year} {2021})}\BibitemShut
  {NoStop}%
\bibitem [{\citenamefont {Larsen}\ \emph {et~al.}(2015)\citenamefont {Larsen},
  \citenamefont {Petersson}, \citenamefont {Kuemmeth}, \citenamefont
  {Jespersen}, \citenamefont {Krogstrup}, \citenamefont {Nyg\aa{}rd},\ and\
  \citenamefont {Marcus}}]{Larsen2015}%
  \BibitemOpen
  \bibfield  {author} {\bibinfo {author} {\bibfnamefont {T.~W.}\ \bibnamefont
  {Larsen}}, \bibinfo {author} {\bibfnamefont {K.~D.}\ \bibnamefont
  {Petersson}}, \bibinfo {author} {\bibfnamefont {F.}~\bibnamefont {Kuemmeth}},
  \bibinfo {author} {\bibfnamefont {T.~S.}\ \bibnamefont {Jespersen}}, \bibinfo
  {author} {\bibfnamefont {P.}~\bibnamefont {Krogstrup}}, \bibinfo {author}
  {\bibfnamefont {J.}~\bibnamefont {Nyg\aa{}rd}},\ and\ \bibinfo {author}
  {\bibfnamefont {C.~M.}\ \bibnamefont {Marcus}},\ }\href
  {https://doi.org/10.1103/PhysRevLett.115.127001} {\bibfield  {journal}
  {\bibinfo  {journal} {Phys. Rev. Lett.}\ }\textbf {\bibinfo {volume} {115}},\
  \bibinfo {pages} {127001} (\bibinfo {year} {2015})}\BibitemShut {NoStop}%
\bibitem [{\citenamefont {Casparis}\ \emph {et~al.}(2018)\citenamefont
  {Casparis}, \citenamefont {Connolly}, \citenamefont {Kjaergaard},
  \citenamefont {Pearson}, \citenamefont {Kringh{\o}j}, \citenamefont {Larsen},
  \citenamefont {Kuemmeth}, \citenamefont {Wang}, \citenamefont {Thomas},
  \citenamefont {Gronin}, \citenamefont {Gardner}, \citenamefont {Manfra},
  \citenamefont {Marcus},\ and\ \citenamefont {Petersson}}]{Casparis2018}%
  \BibitemOpen
  \bibfield  {author} {\bibinfo {author} {\bibfnamefont {L.}~\bibnamefont
  {Casparis}}, \bibinfo {author} {\bibfnamefont {M.~R.}\ \bibnamefont
  {Connolly}}, \bibinfo {author} {\bibfnamefont {M.}~\bibnamefont
  {Kjaergaard}}, \bibinfo {author} {\bibfnamefont {N.~J.}\ \bibnamefont
  {Pearson}}, \bibinfo {author} {\bibfnamefont {A.}~\bibnamefont
  {Kringh{\o}j}}, \bibinfo {author} {\bibfnamefont {T.~W.}\ \bibnamefont
  {Larsen}}, \bibinfo {author} {\bibfnamefont {F.}~\bibnamefont {Kuemmeth}},
  \bibinfo {author} {\bibfnamefont {T.}~\bibnamefont {Wang}}, \bibinfo {author}
  {\bibfnamefont {C.}~\bibnamefont {Thomas}}, \bibinfo {author} {\bibfnamefont
  {S.}~\bibnamefont {Gronin}}, \bibinfo {author} {\bibfnamefont {G.~C.}\
  \bibnamefont {Gardner}}, \bibinfo {author} {\bibfnamefont {M.~J.}\
  \bibnamefont {Manfra}}, \bibinfo {author} {\bibfnamefont {C.~M.}\
  \bibnamefont {Marcus}},\ and\ \bibinfo {author} {\bibfnamefont {K.~D.}\
  \bibnamefont {Petersson}},\ }\href
  {https://doi.org/10.1038/s41565-018-0207-y} {\bibfield  {journal} {\bibinfo
  {journal} {Nature Nanotechnology}\ }\textbf {\bibinfo {volume} {13}},\
  \bibinfo {pages} {915} (\bibinfo {year} {2018})}\BibitemShut {NoStop}%
\bibitem [{\citenamefont {Janvier}\ \emph {et~al.}(2015)\citenamefont
  {Janvier}, \citenamefont {Tosi}, \citenamefont {Bretheau}, \citenamefont
  {Girit}, \citenamefont {Stern}, \citenamefont {Bertet}, \citenamefont
  {Joyez}, \citenamefont {Vion}, \citenamefont {Esteve}, \citenamefont
  {Goffman} \emph {et~al.}}]{Janvier2015}%
  \BibitemOpen
  \bibfield  {author} {\bibinfo {author} {\bibfnamefont {C.}~\bibnamefont
  {Janvier}}, \bibinfo {author} {\bibfnamefont {L.}~\bibnamefont {Tosi}},
  \bibinfo {author} {\bibfnamefont {L.}~\bibnamefont {Bretheau}}, \bibinfo
  {author} {\bibfnamefont {{\c{C}}.}~\bibnamefont {Girit}}, \bibinfo {author}
  {\bibfnamefont {M.}~\bibnamefont {Stern}}, \bibinfo {author} {\bibfnamefont
  {P.}~\bibnamefont {Bertet}}, \bibinfo {author} {\bibfnamefont
  {P.}~\bibnamefont {Joyez}}, \bibinfo {author} {\bibfnamefont
  {D.}~\bibnamefont {Vion}}, \bibinfo {author} {\bibfnamefont {D.}~\bibnamefont
  {Esteve}}, \bibinfo {author} {\bibfnamefont {M.}~\bibnamefont {Goffman}},
  \emph {et~al.},\ }\href@noop {} {\bibfield  {journal} {\bibinfo  {journal}
  {Science}\ }\textbf {\bibinfo {volume} {349}},\ \bibinfo {pages} {1199}
  (\bibinfo {year} {2015})}\BibitemShut {NoStop}%
\bibitem [{\citenamefont {de~Lange}\ \emph {et~al.}(2015)\citenamefont
  {de~Lange}, \citenamefont {van Heck}, \citenamefont {Bruno}, \citenamefont
  {van Woerkom}, \citenamefont {Geresdi}, \citenamefont {Plissard},
  \citenamefont {Bakkers}, \citenamefont {Akhmerov},\ and\ \citenamefont
  {DiCarlo}}]{deLange2015}%
  \BibitemOpen
  \bibfield  {author} {\bibinfo {author} {\bibfnamefont {G.}~\bibnamefont
  {de~Lange}}, \bibinfo {author} {\bibfnamefont {B.}~\bibnamefont {van Heck}},
  \bibinfo {author} {\bibfnamefont {A.}~\bibnamefont {Bruno}}, \bibinfo
  {author} {\bibfnamefont {D.~J.}\ \bibnamefont {van Woerkom}}, \bibinfo
  {author} {\bibfnamefont {A.}~\bibnamefont {Geresdi}}, \bibinfo {author}
  {\bibfnamefont {S.~R.}\ \bibnamefont {Plissard}}, \bibinfo {author}
  {\bibfnamefont {E.~P. A.~M.}\ \bibnamefont {Bakkers}}, \bibinfo {author}
  {\bibfnamefont {A.~R.}\ \bibnamefont {Akhmerov}},\ and\ \bibinfo {author}
  {\bibfnamefont {L.}~\bibnamefont {DiCarlo}},\ }\href
  {https://doi.org/10.1103/PhysRevLett.115.127002} {\bibfield  {journal}
  {\bibinfo  {journal} {Phys. Rev. Lett.}\ }\textbf {\bibinfo {volume} {115}},\
  \bibinfo {pages} {127002} (\bibinfo {year} {2015})}\BibitemShut {NoStop}%
\bibitem [{\citenamefont {Pita-Vidal}\ \emph {et~al.}(2020)\citenamefont
  {Pita-Vidal}, \citenamefont {Bargerbos}, \citenamefont {Yang}, \citenamefont
  {van Woerkom}, \citenamefont {Pfaff}, \citenamefont {Haider}, \citenamefont
  {Krogstrup}, \citenamefont {Kouwenhoven}, \citenamefont {de~Lange},\ and\
  \citenamefont {Kou}}]{Vidal2020}%
  \BibitemOpen
  \bibfield  {author} {\bibinfo {author} {\bibfnamefont {M.}~\bibnamefont
  {Pita-Vidal}}, \bibinfo {author} {\bibfnamefont {A.}~\bibnamefont
  {Bargerbos}}, \bibinfo {author} {\bibfnamefont {C.-K.}\ \bibnamefont {Yang}},
  \bibinfo {author} {\bibfnamefont {D.~J.}\ \bibnamefont {van Woerkom}},
  \bibinfo {author} {\bibfnamefont {W.}~\bibnamefont {Pfaff}}, \bibinfo
  {author} {\bibfnamefont {N.}~\bibnamefont {Haider}}, \bibinfo {author}
  {\bibfnamefont {P.}~\bibnamefont {Krogstrup}}, \bibinfo {author}
  {\bibfnamefont {L.~P.}\ \bibnamefont {Kouwenhoven}}, \bibinfo {author}
  {\bibfnamefont {G.}~\bibnamefont {de~Lange}},\ and\ \bibinfo {author}
  {\bibfnamefont {A.}~\bibnamefont {Kou}},\ }\href
  {https://doi.org/10.1103/PhysRevApplied.14.064038} {\bibfield  {journal}
  {\bibinfo  {journal} {Phys. Rev. Applied}\ }\textbf {\bibinfo {volume}
  {14}},\ \bibinfo {pages} {064038} (\bibinfo {year} {2020})}\BibitemShut
  {NoStop}%
\bibitem [{\citenamefont {Wang}\ \emph {et~al.}(2019)\citenamefont {Wang},
  \citenamefont {Rodan-Legrain}, \citenamefont {Bretheau}, \citenamefont
  {Campbell}, \citenamefont {Kannan}, \citenamefont {Kim}, \citenamefont
  {Kjaergaard}, \citenamefont {Krantz}, \citenamefont {Samach}, \citenamefont
  {Yan}, \citenamefont {Yoder}, \citenamefont {Watanabe}, \citenamefont
  {Taniguchi}, \citenamefont {Orlando}, \citenamefont {Gustavsson},
  \citenamefont {Jarillo-Herrero},\ and\ \citenamefont {Oliver}}]{Wang2019}%
  \BibitemOpen
  \bibfield  {author} {\bibinfo {author} {\bibfnamefont {J.~I.-J.}\
  \bibnamefont {Wang}}, \bibinfo {author} {\bibfnamefont {D.}~\bibnamefont
  {Rodan-Legrain}}, \bibinfo {author} {\bibfnamefont {L.}~\bibnamefont
  {Bretheau}}, \bibinfo {author} {\bibfnamefont {D.~L.}\ \bibnamefont
  {Campbell}}, \bibinfo {author} {\bibfnamefont {B.}~\bibnamefont {Kannan}},
  \bibinfo {author} {\bibfnamefont {D.}~\bibnamefont {Kim}}, \bibinfo {author}
  {\bibfnamefont {M.}~\bibnamefont {Kjaergaard}}, \bibinfo {author}
  {\bibfnamefont {P.}~\bibnamefont {Krantz}}, \bibinfo {author} {\bibfnamefont
  {G.~O.}\ \bibnamefont {Samach}}, \bibinfo {author} {\bibfnamefont
  {F.}~\bibnamefont {Yan}}, \bibinfo {author} {\bibfnamefont {J.~L.}\
  \bibnamefont {Yoder}}, \bibinfo {author} {\bibfnamefont {K.}~\bibnamefont
  {Watanabe}}, \bibinfo {author} {\bibfnamefont {T.}~\bibnamefont {Taniguchi}},
  \bibinfo {author} {\bibfnamefont {T.~P.}\ \bibnamefont {Orlando}}, \bibinfo
  {author} {\bibfnamefont {S.}~\bibnamefont {Gustavsson}}, \bibinfo {author}
  {\bibfnamefont {P.}~\bibnamefont {Jarillo-Herrero}},\ and\ \bibinfo {author}
  {\bibfnamefont {W.~D.}\ \bibnamefont {Oliver}},\ }\href
  {https://doi.org/10.1038/s41565-018-0329-2} {\bibfield  {journal} {\bibinfo
  {journal} {Nature Nanotechnology}\ }\textbf {\bibinfo {volume} {14}},\
  \bibinfo {pages} {120} (\bibinfo {year} {2019})}\BibitemShut {NoStop}%
\bibitem [{\citenamefont {Hays}\ \emph {et~al.}(2021)\citenamefont {Hays},
  \citenamefont {Fatemi}, \citenamefont {Bouman}, \citenamefont {Cerrillo},
  \citenamefont {Diamond}, \citenamefont {Serniak}, \citenamefont {Connolly},
  \citenamefont {Krogstrup}, \citenamefont {Nygård}, \citenamefont {Yeyati},
  \citenamefont {Geresdi},\ and\ \citenamefont {Devoret}}]{Hays2021}%
  \BibitemOpen
  \bibfield  {author} {\bibinfo {author} {\bibfnamefont {M.}~\bibnamefont
  {Hays}}, \bibinfo {author} {\bibfnamefont {V.}~\bibnamefont {Fatemi}},
  \bibinfo {author} {\bibfnamefont {D.}~\bibnamefont {Bouman}}, \bibinfo
  {author} {\bibfnamefont {J.}~\bibnamefont {Cerrillo}}, \bibinfo {author}
  {\bibfnamefont {S.}~\bibnamefont {Diamond}}, \bibinfo {author} {\bibfnamefont
  {K.}~\bibnamefont {Serniak}}, \bibinfo {author} {\bibfnamefont
  {T.}~\bibnamefont {Connolly}}, \bibinfo {author} {\bibfnamefont
  {P.}~\bibnamefont {Krogstrup}}, \bibinfo {author} {\bibfnamefont
  {J.}~\bibnamefont {Nygård}}, \bibinfo {author} {\bibfnamefont {A.~L.}\
  \bibnamefont {Yeyati}}, \bibinfo {author} {\bibfnamefont {A.}~\bibnamefont
  {Geresdi}},\ and\ \bibinfo {author} {\bibfnamefont {M.~H.}\ \bibnamefont
  {Devoret}},\ }\href {https://doi.org/10.1126/science.abf0345} {\bibfield
  {journal} {\bibinfo  {journal} {Science}\ }\textbf {\bibinfo {volume}
  {373}},\ \bibinfo {pages} {430} (\bibinfo {year} {2021})}\BibitemShut
  {NoStop}%
\bibitem [{\citenamefont {Hart}\ \emph {et~al.}(2017)\citenamefont {Hart},
  \citenamefont {Ren}, \citenamefont {Kosowsky}, \citenamefont {Ben-Shach},
  \citenamefont {Leubner}, \citenamefont {Brüne}, \citenamefont {Buhmann},
  \citenamefont {Molenkamp}, \citenamefont {Halperin},\ and\ \citenamefont
  {Yacoby}}]{Hart2017}%
  \BibitemOpen
  \bibfield  {author} {\bibinfo {author} {\bibfnamefont {S.}~\bibnamefont
  {Hart}}, \bibinfo {author} {\bibfnamefont {H.}~\bibnamefont {Ren}}, \bibinfo
  {author} {\bibfnamefont {M.}~\bibnamefont {Kosowsky}}, \bibinfo {author}
  {\bibfnamefont {G.}~\bibnamefont {Ben-Shach}}, \bibinfo {author}
  {\bibfnamefont {P.}~\bibnamefont {Leubner}}, \bibinfo {author} {\bibfnamefont
  {C.}~\bibnamefont {Brüne}}, \bibinfo {author} {\bibfnamefont
  {H.}~\bibnamefont {Buhmann}}, \bibinfo {author} {\bibfnamefont {L.~W.}\
  \bibnamefont {Molenkamp}}, \bibinfo {author} {\bibfnamefont {B.~I.}\
  \bibnamefont {Halperin}},\ and\ \bibinfo {author} {\bibfnamefont
  {A.}~\bibnamefont {Yacoby}},\ }\href {https://doi.org/10.1038/nphys3877}
  {\bibfield  {journal} {\bibinfo  {journal} {Nature Physics}\ }\textbf
  {\bibinfo {volume} {13}},\ \bibinfo {pages} {87} (\bibinfo {year}
  {2017})}\BibitemShut {NoStop}%
\bibitem [{\citenamefont {Mayer}\ \emph {et~al.}(2020)\citenamefont {Mayer},
  \citenamefont {Dartiailh}, \citenamefont {Yuan}, \citenamefont
  {Wickramasinghe}, \citenamefont {Rossi},\ and\ \citenamefont
  {Shabani}}]{Mayer2020}%
  \BibitemOpen
  \bibfield  {author} {\bibinfo {author} {\bibfnamefont {W.}~\bibnamefont
  {Mayer}}, \bibinfo {author} {\bibfnamefont {M.~C.}\ \bibnamefont
  {Dartiailh}}, \bibinfo {author} {\bibfnamefont {J.}~\bibnamefont {Yuan}},
  \bibinfo {author} {\bibfnamefont {K.~S.}\ \bibnamefont {Wickramasinghe}},
  \bibinfo {author} {\bibfnamefont {E.}~\bibnamefont {Rossi}},\ and\ \bibinfo
  {author} {\bibfnamefont {J.}~\bibnamefont {Shabani}},\ }\href
  {https://doi.org/10.1038/s41467-019-14094-1} {\bibfield  {journal} {\bibinfo
  {journal} {Nature Communications}\ }\textbf {\bibinfo {volume} {11}},\
  \bibinfo {pages} {212} (\bibinfo {year} {2020})}\BibitemShut {NoStop}%
\bibitem [{\citenamefont {Baumgartner}\ \emph {et~al.}(2022)\citenamefont
  {Baumgartner}, \citenamefont {Fuchs}, \citenamefont {Costa}, \citenamefont
  {Reinhardt}, \citenamefont {Gronin}, \citenamefont {Gardner}, \citenamefont
  {Lindemann}, \citenamefont {Manfra}, \citenamefont {{Faria Junior}},
  \citenamefont {Kochan}, \citenamefont {Fabian}, \citenamefont {Paradiso},\
  and\ \citenamefont {Strunk}}]{Baumgartner2022}%
  \BibitemOpen
  \bibfield  {author} {\bibinfo {author} {\bibfnamefont {C.}~\bibnamefont
  {Baumgartner}}, \bibinfo {author} {\bibfnamefont {L.}~\bibnamefont {Fuchs}},
  \bibinfo {author} {\bibfnamefont {A.}~\bibnamefont {Costa}}, \bibinfo
  {author} {\bibfnamefont {S.}~\bibnamefont {Reinhardt}}, \bibinfo {author}
  {\bibfnamefont {S.}~\bibnamefont {Gronin}}, \bibinfo {author} {\bibfnamefont
  {G.~C.}\ \bibnamefont {Gardner}}, \bibinfo {author} {\bibfnamefont
  {T.}~\bibnamefont {Lindemann}}, \bibinfo {author} {\bibfnamefont {M.~J.}\
  \bibnamefont {Manfra}}, \bibinfo {author} {\bibfnamefont {P.~E.}\
  \bibnamefont {{Faria Junior}}}, \bibinfo {author} {\bibfnamefont
  {D.}~\bibnamefont {Kochan}}, \bibinfo {author} {\bibfnamefont
  {J.}~\bibnamefont {Fabian}}, \bibinfo {author} {\bibfnamefont
  {N.}~\bibnamefont {Paradiso}},\ and\ \bibinfo {author} {\bibfnamefont
  {C.}~\bibnamefont {Strunk}},\ }\href
  {https://doi.org/10.1038/s41565-021-01009-9} {\bibfield  {journal} {\bibinfo
  {journal} {Nature Nanotechnology}\ }\textbf {\bibinfo {volume} {17}},\
  \bibinfo {pages} {39} (\bibinfo {year} {2022})}\BibitemShut {NoStop}%
\bibitem [{\citenamefont {Hell}\ \emph {et~al.}(2017)\citenamefont {Hell},
  \citenamefont {Leijnse},\ and\ \citenamefont {Flensberg}}]{Hell2017}%
  \BibitemOpen
  \bibfield  {author} {\bibinfo {author} {\bibfnamefont {M.}~\bibnamefont
  {Hell}}, \bibinfo {author} {\bibfnamefont {M.}~\bibnamefont {Leijnse}},\ and\
  \bibinfo {author} {\bibfnamefont {K.}~\bibnamefont {Flensberg}},\ }\href
  {https://doi.org/10.1103/PhysRevLett.118.107701} {\bibfield  {journal}
  {\bibinfo  {journal} {Phys. Rev. Lett.}\ }\textbf {\bibinfo {volume} {118}},\
  \bibinfo {pages} {107701} (\bibinfo {year} {2017})}\BibitemShut {NoStop}%
\bibitem [{\citenamefont {Pientka}\ \emph {et~al.}(2017)\citenamefont
  {Pientka}, \citenamefont {Keselman}, \citenamefont {Berg}, \citenamefont
  {Yacoby}, \citenamefont {Stern},\ and\ \citenamefont
  {Halperin}}]{Pientka2017}%
  \BibitemOpen
  \bibfield  {author} {\bibinfo {author} {\bibfnamefont {F.}~\bibnamefont
  {Pientka}}, \bibinfo {author} {\bibfnamefont {A.}~\bibnamefont {Keselman}},
  \bibinfo {author} {\bibfnamefont {E.}~\bibnamefont {Berg}}, \bibinfo {author}
  {\bibfnamefont {A.}~\bibnamefont {Yacoby}}, \bibinfo {author} {\bibfnamefont
  {A.}~\bibnamefont {Stern}},\ and\ \bibinfo {author} {\bibfnamefont {B.~I.}\
  \bibnamefont {Halperin}},\ }\href {https://doi.org/10.1103/PhysRevX.7.021032}
  {\bibfield  {journal} {\bibinfo  {journal} {Phys. Rev. X}\ }\textbf {\bibinfo
  {volume} {7}},\ \bibinfo {pages} {021032} (\bibinfo {year}
  {2017})}\BibitemShut {NoStop}%
\bibitem [{\citenamefont {Fornieri}\ \emph {et~al.}(2019)\citenamefont
  {Fornieri}, \citenamefont {Whiticar}, \citenamefont {Setiawan}, \citenamefont
  {Portol{\'{e}}s}, \citenamefont {Drachmann}, \citenamefont {Keselman},
  \citenamefont {Gronin}, \citenamefont {Thomas}, \citenamefont {Wang},
  \citenamefont {Kallaher}, \citenamefont {Gardner}, \citenamefont {Berg},
  \citenamefont {Manfra}, \citenamefont {Stern}, \citenamefont {Marcus},\ and\
  \citenamefont {Nichele}}]{Fornieri2019}%
  \BibitemOpen
  \bibfield  {author} {\bibinfo {author} {\bibfnamefont {A.}~\bibnamefont
  {Fornieri}}, \bibinfo {author} {\bibfnamefont {A.~M.}\ \bibnamefont
  {Whiticar}}, \bibinfo {author} {\bibfnamefont {F.}~\bibnamefont {Setiawan}},
  \bibinfo {author} {\bibfnamefont {E.}~\bibnamefont {Portol{\'{e}}s}},
  \bibinfo {author} {\bibfnamefont {A.~C.~C.}\ \bibnamefont {Drachmann}},
  \bibinfo {author} {\bibfnamefont {A.}~\bibnamefont {Keselman}}, \bibinfo
  {author} {\bibfnamefont {S.}~\bibnamefont {Gronin}}, \bibinfo {author}
  {\bibfnamefont {C.}~\bibnamefont {Thomas}}, \bibinfo {author} {\bibfnamefont
  {T.}~\bibnamefont {Wang}}, \bibinfo {author} {\bibfnamefont {R.}~\bibnamefont
  {Kallaher}}, \bibinfo {author} {\bibfnamefont {G.~C.}\ \bibnamefont
  {Gardner}}, \bibinfo {author} {\bibfnamefont {E.}~\bibnamefont {Berg}},
  \bibinfo {author} {\bibfnamefont {M.~J.}\ \bibnamefont {Manfra}}, \bibinfo
  {author} {\bibfnamefont {A.}~\bibnamefont {Stern}}, \bibinfo {author}
  {\bibfnamefont {C.~M.}\ \bibnamefont {Marcus}},\ and\ \bibinfo {author}
  {\bibfnamefont {F.}~\bibnamefont {Nichele}},\ }\href
  {https://doi.org/10.1038/s41586-019-1068-8} {\bibfield  {journal} {\bibinfo
  {journal} {Nature}\ }\textbf {\bibinfo {volume} {569}},\ \bibinfo {pages}
  {89} (\bibinfo {year} {2019})}\BibitemShut {NoStop}%
\bibitem [{\citenamefont {Ren}\ \emph {et~al.}(2019)\citenamefont {Ren},
  \citenamefont {Pientka}, \citenamefont {Hart}, \citenamefont {Pierce},
  \citenamefont {Kosowsky}, \citenamefont {Lunczer}, \citenamefont {Schlereth},
  \citenamefont {Scharf}, \citenamefont {Hankiewicz}, \citenamefont
  {Molenkamp}, \citenamefont {Halperin},\ and\ \citenamefont
  {Yacoby}}]{Ren2019}%
  \BibitemOpen
  \bibfield  {author} {\bibinfo {author} {\bibfnamefont {H.}~\bibnamefont
  {Ren}}, \bibinfo {author} {\bibfnamefont {F.}~\bibnamefont {Pientka}},
  \bibinfo {author} {\bibfnamefont {S.}~\bibnamefont {Hart}}, \bibinfo {author}
  {\bibfnamefont {A.~T.}\ \bibnamefont {Pierce}}, \bibinfo {author}
  {\bibfnamefont {M.}~\bibnamefont {Kosowsky}}, \bibinfo {author}
  {\bibfnamefont {L.}~\bibnamefont {Lunczer}}, \bibinfo {author} {\bibfnamefont
  {R.}~\bibnamefont {Schlereth}}, \bibinfo {author} {\bibfnamefont
  {B.}~\bibnamefont {Scharf}}, \bibinfo {author} {\bibfnamefont {E.~M.}\
  \bibnamefont {Hankiewicz}}, \bibinfo {author} {\bibfnamefont {L.~W.}\
  \bibnamefont {Molenkamp}}, \bibinfo {author} {\bibfnamefont {B.~I.}\
  \bibnamefont {Halperin}},\ and\ \bibinfo {author} {\bibfnamefont
  {A.}~\bibnamefont {Yacoby}},\ }\href
  {https://doi.org/10.1038/s41586-019-1148-9} {\bibfield  {journal} {\bibinfo
  {journal} {Nature}\ }\textbf {\bibinfo {volume} {569}},\ \bibinfo {pages}
  {93} (\bibinfo {year} {2019})}\BibitemShut {NoStop}%
\bibitem [{\citenamefont {Dartiailh}\ \emph {et~al.}(2021)\citenamefont
  {Dartiailh}, \citenamefont {Mayer}, \citenamefont {Yuan}, \citenamefont
  {Wickramasinghe}, \citenamefont {Matos-Abiague}, \citenamefont {\ifmmode
  \check{Z}\else \v{Z}\fi{}uti\ifmmode~\acute{c}\else \'{c}\fi{}},\ and\
  \citenamefont {Shabani}}]{Dartiailh2021}%
  \BibitemOpen
  \bibfield  {author} {\bibinfo {author} {\bibfnamefont {M.~C.}\ \bibnamefont
  {Dartiailh}}, \bibinfo {author} {\bibfnamefont {W.}~\bibnamefont {Mayer}},
  \bibinfo {author} {\bibfnamefont {J.}~\bibnamefont {Yuan}}, \bibinfo {author}
  {\bibfnamefont {K.~S.}\ \bibnamefont {Wickramasinghe}}, \bibinfo {author}
  {\bibfnamefont {A.}~\bibnamefont {Matos-Abiague}}, \bibinfo {author}
  {\bibfnamefont {I.}~\bibnamefont {\ifmmode \check{Z}\else
  \v{Z}\fi{}uti\ifmmode~\acute{c}\else \'{c}\fi{}}},\ and\ \bibinfo {author}
  {\bibfnamefont {J.}~\bibnamefont {Shabani}},\ }\href
  {https://doi.org/10.1103/PhysRevLett.126.036802} {\bibfield  {journal}
  {\bibinfo  {journal} {Phys. Rev. Lett.}\ }\textbf {\bibinfo {volume} {126}},\
  \bibinfo {pages} {036802} (\bibinfo {year} {2021})}\BibitemShut {NoStop}%
\bibitem [{\citenamefont {Banerjee}\ \emph {et~al.}(2022)\citenamefont
  {Banerjee}, \citenamefont {Lesser}, \citenamefont {Rahman}, \citenamefont
  {Wang}, \citenamefont {Li}, \citenamefont {Kringhøj}, \citenamefont
  {Whiticar}, \citenamefont {Drachmann}, \citenamefont {Thomas}, \citenamefont
  {Wang}, \citenamefont {Manfra}, \citenamefont {Berg}, \citenamefont {Oreg},
  \citenamefont {Stern},\ and\ \citenamefont {Marcus}}]{Banerjee2022}%
  \BibitemOpen
  \bibfield  {author} {\bibinfo {author} {\bibfnamefont {A.}~\bibnamefont
  {Banerjee}}, \bibinfo {author} {\bibfnamefont {O.}~\bibnamefont {Lesser}},
  \bibinfo {author} {\bibfnamefont {M.~A.}\ \bibnamefont {Rahman}}, \bibinfo
  {author} {\bibfnamefont {H.~R.}\ \bibnamefont {Wang}}, \bibinfo {author}
  {\bibfnamefont {M.~R.}\ \bibnamefont {Li}}, \bibinfo {author} {\bibfnamefont
  {A.}~\bibnamefont {Kringhøj}}, \bibinfo {author} {\bibfnamefont {A.~M.}\
  \bibnamefont {Whiticar}}, \bibinfo {author} {\bibfnamefont {A.~C.~C.}\
  \bibnamefont {Drachmann}}, \bibinfo {author} {\bibfnamefont {C.}~\bibnamefont
  {Thomas}}, \bibinfo {author} {\bibfnamefont {T.}~\bibnamefont {Wang}},
  \bibinfo {author} {\bibfnamefont {M.~J.}\ \bibnamefont {Manfra}}, \bibinfo
  {author} {\bibfnamefont {E.}~\bibnamefont {Berg}}, \bibinfo {author}
  {\bibfnamefont {Y.}~\bibnamefont {Oreg}}, \bibinfo {author} {\bibfnamefont
  {A.}~\bibnamefont {Stern}},\ and\ \bibinfo {author} {\bibfnamefont {C.~M.}\
  \bibnamefont {Marcus}},\ }\href@noop {} {\bibinfo {title} {Signatures of a
  topological phase transition in a planar josephson junction}} (\bibinfo
  {year} {2022}),\ \Eprint {https://arxiv.org/abs/2201.03453} {arXiv:2201.03453
  [cond-mat.mes-hall]} \BibitemShut {NoStop}%
\bibitem [{\citenamefont {Nichele}\ \emph {et~al.}(2020)\citenamefont
  {Nichele}, \citenamefont {Portol\'es}, \citenamefont {Fornieri},
  \citenamefont {Whiticar}, \citenamefont {Drachmann}, \citenamefont {Gronin},
  \citenamefont {Wang}, \citenamefont {Gardner}, \citenamefont {Thomas},
  \citenamefont {Hatke}, \citenamefont {Manfra},\ and\ \citenamefont
  {Marcus}}]{Nichele2020}%
  \BibitemOpen
  \bibfield  {author} {\bibinfo {author} {\bibfnamefont {F.}~\bibnamefont
  {Nichele}}, \bibinfo {author} {\bibfnamefont {E.}~\bibnamefont {Portol\'es}},
  \bibinfo {author} {\bibfnamefont {A.}~\bibnamefont {Fornieri}}, \bibinfo
  {author} {\bibfnamefont {A.~M.}\ \bibnamefont {Whiticar}}, \bibinfo {author}
  {\bibfnamefont {A.~C.~C.}\ \bibnamefont {Drachmann}}, \bibinfo {author}
  {\bibfnamefont {S.}~\bibnamefont {Gronin}}, \bibinfo {author} {\bibfnamefont
  {T.}~\bibnamefont {Wang}}, \bibinfo {author} {\bibfnamefont {G.~C.}\
  \bibnamefont {Gardner}}, \bibinfo {author} {\bibfnamefont {C.}~\bibnamefont
  {Thomas}}, \bibinfo {author} {\bibfnamefont {A.~T.}\ \bibnamefont {Hatke}},
  \bibinfo {author} {\bibfnamefont {M.~J.}\ \bibnamefont {Manfra}},\ and\
  \bibinfo {author} {\bibfnamefont {C.~M.}\ \bibnamefont {Marcus}},\ }\href
  {https://doi.org/10.1103/PhysRevLett.124.226801} {\bibfield  {journal}
  {\bibinfo  {journal} {Phys. Rev. Lett.}\ }\textbf {\bibinfo {volume} {124}},\
  \bibinfo {pages} {226801} (\bibinfo {year} {2020})}\BibitemShut {NoStop}%
\bibitem [{\citenamefont {Suominen}\ \emph {et~al.}(2017)\citenamefont
  {Suominen}, \citenamefont {Danon}, \citenamefont {Kjaergaard}, \citenamefont
  {Flensberg}, \citenamefont {Shabani}, \citenamefont {Palmstr\o{}m},
  \citenamefont {Nichele},\ and\ \citenamefont {Marcus}}]{Suominen2017}%
  \BibitemOpen
  \bibfield  {author} {\bibinfo {author} {\bibfnamefont {H.~J.}\ \bibnamefont
  {Suominen}}, \bibinfo {author} {\bibfnamefont {J.}~\bibnamefont {Danon}},
  \bibinfo {author} {\bibfnamefont {M.}~\bibnamefont {Kjaergaard}}, \bibinfo
  {author} {\bibfnamefont {K.}~\bibnamefont {Flensberg}}, \bibinfo {author}
  {\bibfnamefont {J.}~\bibnamefont {Shabani}}, \bibinfo {author} {\bibfnamefont
  {C.~J.}\ \bibnamefont {Palmstr\o{}m}}, \bibinfo {author} {\bibfnamefont
  {F.}~\bibnamefont {Nichele}},\ and\ \bibinfo {author} {\bibfnamefont {C.~M.}\
  \bibnamefont {Marcus}},\ }\href {https://doi.org/10.1103/PhysRevB.95.035307}
  {\bibfield  {journal} {\bibinfo  {journal} {Phys. Rev. B}\ }\textbf {\bibinfo
  {volume} {95}},\ \bibinfo {pages} {035307} (\bibinfo {year}
  {2017})}\BibitemShut {NoStop}%
\bibitem [{\citenamefont {Fulton}\ and\ \citenamefont
  {Dunkleberger}(1974)}]{Fulton1974}%
  \BibitemOpen
  \bibfield  {author} {\bibinfo {author} {\bibfnamefont {T.~A.}\ \bibnamefont
  {Fulton}}\ and\ \bibinfo {author} {\bibfnamefont {L.~N.}\ \bibnamefont
  {Dunkleberger}},\ }\href {https://doi.org/10.1103/PhysRevB.9.4760} {\bibfield
   {journal} {\bibinfo  {journal} {Phys. Rev. B}\ }\textbf {\bibinfo {volume}
  {9}},\ \bibinfo {pages} {4760} (\bibinfo {year} {1974})}\BibitemShut
  {NoStop}%
\bibitem [{\citenamefont {Martinis}\ \emph {et~al.}(1987)\citenamefont
  {Martinis}, \citenamefont {Devoret},\ and\ \citenamefont
  {Clarke}}]{Martinis1987}%
  \BibitemOpen
  \bibfield  {author} {\bibinfo {author} {\bibfnamefont {J.~M.}\ \bibnamefont
  {Martinis}}, \bibinfo {author} {\bibfnamefont {M.~H.}\ \bibnamefont
  {Devoret}},\ and\ \bibinfo {author} {\bibfnamefont {J.}~\bibnamefont
  {Clarke}},\ }\href {https://doi.org/10.1103/PhysRevB.35.4682} {\bibfield
  {journal} {\bibinfo  {journal} {Phys. Rev. B}\ }\textbf {\bibinfo {volume}
  {35}},\ \bibinfo {pages} {4682} (\bibinfo {year} {1987})}\BibitemShut
  {NoStop}%
\bibitem [{\citenamefont {Lee}\ \emph {et~al.}(2011)\citenamefont {Lee},
  \citenamefont {Jeong}, \citenamefont {Choi}, \citenamefont {Doh},\ and\
  \citenamefont {Lee}}]{Lee2011}%
  \BibitemOpen
  \bibfield  {author} {\bibinfo {author} {\bibfnamefont {G.-H.}\ \bibnamefont
  {Lee}}, \bibinfo {author} {\bibfnamefont {D.}~\bibnamefont {Jeong}}, \bibinfo
  {author} {\bibfnamefont {J.-H.}\ \bibnamefont {Choi}}, \bibinfo {author}
  {\bibfnamefont {Y.-J.}\ \bibnamefont {Doh}},\ and\ \bibinfo {author}
  {\bibfnamefont {H.-J.}\ \bibnamefont {Lee}},\ }\href
  {https://doi.org/10.1103/PhysRevLett.107.146605} {\bibfield  {journal}
  {\bibinfo  {journal} {Phys. Rev. Lett.}\ }\textbf {\bibinfo {volume} {107}},\
  \bibinfo {pages} {146605} (\bibinfo {year} {2011})}\BibitemShut {NoStop}%
\bibitem [{\citenamefont {Murphy}\ \emph {et~al.}(2013)\citenamefont {Murphy},
  \citenamefont {Weinberg}, \citenamefont {Aref}, \citenamefont {Coskun},
  \citenamefont {Vakaryuk}, \citenamefont {Levchenko},\ and\ \citenamefont
  {Bezryadin}}]{Murphy2013}%
  \BibitemOpen
  \bibfield  {author} {\bibinfo {author} {\bibfnamefont {A.}~\bibnamefont
  {Murphy}}, \bibinfo {author} {\bibfnamefont {P.}~\bibnamefont {Weinberg}},
  \bibinfo {author} {\bibfnamefont {T.}~\bibnamefont {Aref}}, \bibinfo {author}
  {\bibfnamefont {U.~C.}\ \bibnamefont {Coskun}}, \bibinfo {author}
  {\bibfnamefont {V.}~\bibnamefont {Vakaryuk}}, \bibinfo {author}
  {\bibfnamefont {A.}~\bibnamefont {Levchenko}},\ and\ \bibinfo {author}
  {\bibfnamefont {A.}~\bibnamefont {Bezryadin}},\ }\href
  {https://doi.org/10.1103/PhysRevLett.110.247001} {\bibfield  {journal}
  {\bibinfo  {journal} {Phys. Rev. Lett.}\ }\textbf {\bibinfo {volume} {110}},\
  \bibinfo {pages} {247001} (\bibinfo {year} {2013})}\BibitemShut {NoStop}%
\bibitem [{\citenamefont {Kim}\ and\ \citenamefont {Doh}(2016)}]{Kim2016}%
  \BibitemOpen
  \bibfield  {author} {\bibinfo {author} {\bibfnamefont {B.-K.}\ \bibnamefont
  {Kim}}\ and\ \bibinfo {author} {\bibfnamefont {Y.-J.}\ \bibnamefont {Doh}},\
  }\href {https://doi.org/10.3938/jkps.69.349} {\bibfield  {journal} {\bibinfo
  {journal} {Journal of the Korean Physical Society}\ }\textbf {\bibinfo
  {volume} {69}},\ \bibinfo {pages} {349} (\bibinfo {year} {2016})}\BibitemShut
  {NoStop}%
\bibitem [{\citenamefont {Kim}\ \emph {et~al.}(2017)\citenamefont {Kim},
  \citenamefont {Kim}, \citenamefont {Kim}, \citenamefont {Hwang},
  \citenamefont {Kim},\ and\ \citenamefont {Doh}}]{Kim2017}%
  \BibitemOpen
  \bibfield  {author} {\bibinfo {author} {\bibfnamefont {J.}~\bibnamefont
  {Kim}}, \bibinfo {author} {\bibfnamefont {B.-K.}\ \bibnamefont {Kim}},
  \bibinfo {author} {\bibfnamefont {H.-S.}\ \bibnamefont {Kim}}, \bibinfo
  {author} {\bibfnamefont {A.}~\bibnamefont {Hwang}}, \bibinfo {author}
  {\bibfnamefont {B.}~\bibnamefont {Kim}},\ and\ \bibinfo {author}
  {\bibfnamefont {Y.-J.}\ \bibnamefont {Doh}},\ }\href
  {https://doi.org/10.1021/acs.nanolett.7b03571} {\bibfield  {journal}
  {\bibinfo  {journal} {Nano Lett.}\ }\textbf {\bibinfo {volume} {17}},\
  \bibinfo {pages} {6997} (\bibinfo {year} {2017})}\BibitemShut {NoStop}%
\bibitem [{\citenamefont {Sahu}\ \emph {et~al.}(2009)\citenamefont {Sahu},
  \citenamefont {Bae}, \citenamefont {Rogachev}, \citenamefont {Pekker},
  \citenamefont {Wei}, \citenamefont {Shah}, \citenamefont {Goldbart},\ and\
  \citenamefont {Bezryadin}}]{Sahu2009}%
  \BibitemOpen
  \bibfield  {author} {\bibinfo {author} {\bibfnamefont {M.}~\bibnamefont
  {Sahu}}, \bibinfo {author} {\bibfnamefont {M.-H.}\ \bibnamefont {Bae}},
  \bibinfo {author} {\bibfnamefont {A.}~\bibnamefont {Rogachev}}, \bibinfo
  {author} {\bibfnamefont {D.}~\bibnamefont {Pekker}}, \bibinfo {author}
  {\bibfnamefont {T.-C.}\ \bibnamefont {Wei}}, \bibinfo {author} {\bibfnamefont
  {N.}~\bibnamefont {Shah}}, \bibinfo {author} {\bibfnamefont {P.~M.}\
  \bibnamefont {Goldbart}},\ and\ \bibinfo {author} {\bibfnamefont
  {A.}~\bibnamefont {Bezryadin}},\ }\href {https://doi.org/10.1038/nphys1276}
  {\bibfield  {journal} {\bibinfo  {journal} {Nature Physics}\ }\textbf
  {\bibinfo {volume} {5}},\ \bibinfo {pages} {503} (\bibinfo {year}
  {2009})}\BibitemShut {NoStop}%
\bibitem [{\citenamefont {Li}\ \emph {et~al.}(2011)\citenamefont {Li},
  \citenamefont {Wu}, \citenamefont {Bomze}, \citenamefont {Borzenets},
  \citenamefont {Finkelstein},\ and\ \citenamefont {Chang}}]{Li2011}%
  \BibitemOpen
  \bibfield  {author} {\bibinfo {author} {\bibfnamefont {P.}~\bibnamefont
  {Li}}, \bibinfo {author} {\bibfnamefont {P.~M.}\ \bibnamefont {Wu}}, \bibinfo
  {author} {\bibfnamefont {Y.}~\bibnamefont {Bomze}}, \bibinfo {author}
  {\bibfnamefont {I.~V.}\ \bibnamefont {Borzenets}}, \bibinfo {author}
  {\bibfnamefont {G.}~\bibnamefont {Finkelstein}},\ and\ \bibinfo {author}
  {\bibfnamefont {A.~M.}\ \bibnamefont {Chang}},\ }\href
  {https://doi.org/10.1103/PhysRevLett.107.137004} {\bibfield  {journal}
  {\bibinfo  {journal} {Phys. Rev. Lett.}\ }\textbf {\bibinfo {volume} {107}},\
  \bibinfo {pages} {137004} (\bibinfo {year} {2011})}\BibitemShut {NoStop}%
\bibitem [{\citenamefont {Aref}\ \emph {et~al.}(2012)\citenamefont {Aref},
  \citenamefont {Levchenko}, \citenamefont {Vakaryuk},\ and\ \citenamefont
  {Bezryadin}}]{Aref2012}%
  \BibitemOpen
  \bibfield  {author} {\bibinfo {author} {\bibfnamefont {T.}~\bibnamefont
  {Aref}}, \bibinfo {author} {\bibfnamefont {A.}~\bibnamefont {Levchenko}},
  \bibinfo {author} {\bibfnamefont {V.}~\bibnamefont {Vakaryuk}},\ and\
  \bibinfo {author} {\bibfnamefont {A.}~\bibnamefont {Bezryadin}},\ }\href
  {https://doi.org/10.1103/PhysRevB.86.024507} {\bibfield  {journal} {\bibinfo
  {journal} {Phys. Rev. B}\ }\textbf {\bibinfo {volume} {86}},\ \bibinfo
  {pages} {024507} (\bibinfo {year} {2012})}\BibitemShut {NoStop}%
\bibitem [{\citenamefont {Lefevre-Seguin}\ \emph {et~al.}(1992)\citenamefont
  {Lefevre-Seguin}, \citenamefont {Turlot}, \citenamefont {Urbina},
  \citenamefont {Esteve},\ and\ \citenamefont {Devoret}}]{LefevreSeguin1992}%
  \BibitemOpen
  \bibfield  {author} {\bibinfo {author} {\bibfnamefont {V.}~\bibnamefont
  {Lefevre-Seguin}}, \bibinfo {author} {\bibfnamefont {E.}~\bibnamefont
  {Turlot}}, \bibinfo {author} {\bibfnamefont {C.}~\bibnamefont {Urbina}},
  \bibinfo {author} {\bibfnamefont {D.}~\bibnamefont {Esteve}},\ and\ \bibinfo
  {author} {\bibfnamefont {M.~H.}\ \bibnamefont {Devoret}},\ }\href
  {https://doi.org/10.1103/PhysRevB.46.5507} {\bibfield  {journal} {\bibinfo
  {journal} {Phys. Rev. B}\ }\textbf {\bibinfo {volume} {46}},\ \bibinfo
  {pages} {5507} (\bibinfo {year} {1992})}\BibitemShut {NoStop}%
\bibitem [{\citenamefont {Li}\ \emph {et~al.}(2002)\citenamefont {Li},
  \citenamefont {Yu}, \citenamefont {Zhang}, \citenamefont {Qiu}, \citenamefont
  {Han},\ and\ \citenamefont {Wang}}]{Li2002}%
  \BibitemOpen
  \bibfield  {author} {\bibinfo {author} {\bibfnamefont {S.-X.}\ \bibnamefont
  {Li}}, \bibinfo {author} {\bibfnamefont {Y.}~\bibnamefont {Yu}}, \bibinfo
  {author} {\bibfnamefont {Y.}~\bibnamefont {Zhang}}, \bibinfo {author}
  {\bibfnamefont {W.}~\bibnamefont {Qiu}}, \bibinfo {author} {\bibfnamefont
  {S.}~\bibnamefont {Han}},\ and\ \bibinfo {author} {\bibfnamefont
  {Z.}~\bibnamefont {Wang}},\ }\href
  {https://doi.org/10.1103/PhysRevLett.89.098301} {\bibfield  {journal}
  {\bibinfo  {journal} {Phys. Rev. Lett.}\ }\textbf {\bibinfo {volume} {89}},\
  \bibinfo {pages} {098301} (\bibinfo {year} {2002})}\BibitemShut {NoStop}%
\bibitem [{\citenamefont {Balestro}\ \emph {et~al.}(2003)\citenamefont
  {Balestro}, \citenamefont {Claudon}, \citenamefont {Pekola},\ and\
  \citenamefont {Buisson}}]{Balestro2003}%
  \BibitemOpen
  \bibfield  {author} {\bibinfo {author} {\bibfnamefont {F.}~\bibnamefont
  {Balestro}}, \bibinfo {author} {\bibfnamefont {J.}~\bibnamefont {Claudon}},
  \bibinfo {author} {\bibfnamefont {J.~P.}\ \bibnamefont {Pekola}},\ and\
  \bibinfo {author} {\bibfnamefont {O.}~\bibnamefont {Buisson}},\ }\href
  {https://doi.org/10.1103/PhysRevLett.91.158301} {\bibfield  {journal}
  {\bibinfo  {journal} {Phys. Rev. Lett.}\ }\textbf {\bibinfo {volume} {91}},\
  \bibinfo {pages} {158301} (\bibinfo {year} {2003})}\BibitemShut {NoStop}%
\bibitem [{\citenamefont {Sullivan}\ \emph {et~al.}(2013)\citenamefont
  {Sullivan}, \citenamefont {Dutta}, \citenamefont {Dreyer}, \citenamefont
  {Gubrud}, \citenamefont {Roychowdhury}, \citenamefont {Anderson},
  \citenamefont {Lobb},\ and\ \citenamefont {Wellstood}}]{Sullivan2013}%
  \BibitemOpen
  \bibfield  {author} {\bibinfo {author} {\bibfnamefont {D.~F.}\ \bibnamefont
  {Sullivan}}, \bibinfo {author} {\bibfnamefont {S.~K.}\ \bibnamefont {Dutta}},
  \bibinfo {author} {\bibfnamefont {M.}~\bibnamefont {Dreyer}}, \bibinfo
  {author} {\bibfnamefont {M.~A.}\ \bibnamefont {Gubrud}}, \bibinfo {author}
  {\bibfnamefont {A.}~\bibnamefont {Roychowdhury}}, \bibinfo {author}
  {\bibfnamefont {J.~R.}\ \bibnamefont {Anderson}}, \bibinfo {author}
  {\bibfnamefont {C.~J.}\ \bibnamefont {Lobb}},\ and\ \bibinfo {author}
  {\bibfnamefont {F.~C.}\ \bibnamefont {Wellstood}},\ }\href
  {https://doi.org/10.1063/1.4804057} {\bibfield  {journal} {\bibinfo
  {journal} {Journal of Applied Physics}\ }\textbf {\bibinfo {volume} {113}},\
  \bibinfo {pages} {183905} (\bibinfo {year} {2013})},\ \Eprint
  {https://arxiv.org/abs/https://doi.org/10.1063/1.4804057}
  {https://doi.org/10.1063/1.4804057} \BibitemShut {NoStop}%
\bibitem [{\citenamefont {Butz}\ \emph {et~al.}(2014)\citenamefont {Butz},
  \citenamefont {Feofanov}, \citenamefont {Fedorov}, \citenamefont {Rotzinger},
  \citenamefont {Thomann}, \citenamefont {Mackrodt}, \citenamefont {Dolata},
  \citenamefont {Geshkenbein}, \citenamefont {Blatter},\ and\ \citenamefont
  {Ustinov}}]{Butz2014}%
  \BibitemOpen
  \bibfield  {author} {\bibinfo {author} {\bibfnamefont {S.}~\bibnamefont
  {Butz}}, \bibinfo {author} {\bibfnamefont {A.~K.}\ \bibnamefont {Feofanov}},
  \bibinfo {author} {\bibfnamefont {K.~G.}\ \bibnamefont {Fedorov}}, \bibinfo
  {author} {\bibfnamefont {H.}~\bibnamefont {Rotzinger}}, \bibinfo {author}
  {\bibfnamefont {A.~U.}\ \bibnamefont {Thomann}}, \bibinfo {author}
  {\bibfnamefont {B.}~\bibnamefont {Mackrodt}}, \bibinfo {author}
  {\bibfnamefont {R.}~\bibnamefont {Dolata}}, \bibinfo {author} {\bibfnamefont
  {V.~B.}\ \bibnamefont {Geshkenbein}}, \bibinfo {author} {\bibfnamefont
  {G.}~\bibnamefont {Blatter}},\ and\ \bibinfo {author} {\bibfnamefont {A.~V.}\
  \bibnamefont {Ustinov}},\ }\href
  {https://doi.org/10.1103/PhysRevLett.113.247005} {\bibfield  {journal}
  {\bibinfo  {journal} {Phys. Rev. Lett.}\ }\textbf {\bibinfo {volume} {113}},\
  \bibinfo {pages} {247005} (\bibinfo {year} {2014})}\BibitemShut {NoStop}%
\bibitem [{\citenamefont {Bezryadin}(2012)}]{Bezryadin2012}%
  \BibitemOpen
  \bibfield  {author} {\bibinfo {author} {\bibfnamefont {A.}~\bibnamefont
  {Bezryadin}},\ }\bibinfo {title} {Stochastic premature switching and
  {K}urkij{\"a}rvi theory},\ in\ \href
  {https://doi.org/https://doi.org/10.1002/9783527651931.ch8} {\emph {\bibinfo
  {booktitle} {Superconductivity in Nanowires}}}\ (\bibinfo  {publisher} {John
  Wiley {\&} Sons, Ltd},\ \bibinfo {year} {2012})\ Chap.~\bibinfo {chapter}
  {8}, pp.\ \bibinfo {pages} {131--162}\BibitemShut {NoStop}%
\bibitem [{\citenamefont {Longobardi}\ \emph {et~al.}(2012)\citenamefont
  {Longobardi}, \citenamefont {Massarotti}, \citenamefont {Stornaiuolo},
  \citenamefont {Galletti}, \citenamefont {Rotoli}, \citenamefont {Lombardi},\
  and\ \citenamefont {Tafuri}}]{Longobardi2012}%
  \BibitemOpen
  \bibfield  {author} {\bibinfo {author} {\bibfnamefont {L.}~\bibnamefont
  {Longobardi}}, \bibinfo {author} {\bibfnamefont {D.}~\bibnamefont
  {Massarotti}}, \bibinfo {author} {\bibfnamefont {D.}~\bibnamefont
  {Stornaiuolo}}, \bibinfo {author} {\bibfnamefont {L.}~\bibnamefont
  {Galletti}}, \bibinfo {author} {\bibfnamefont {G.}~\bibnamefont {Rotoli}},
  \bibinfo {author} {\bibfnamefont {F.}~\bibnamefont {Lombardi}},\ and\
  \bibinfo {author} {\bibfnamefont {F.}~\bibnamefont {Tafuri}},\ }\href
  {https://doi.org/10.1103/PhysRevLett.109.050601} {\bibfield  {journal}
  {\bibinfo  {journal} {Phys. Rev. Lett.}\ }\textbf {\bibinfo {volume} {109}},\
  \bibinfo {pages} {050601} (\bibinfo {year} {2012})}\BibitemShut {NoStop}%
\bibitem [{Sup()}]{Supplement}%
  \BibitemOpen
  \href@noop {} {}\bibinfo {note} {See the Supplemental Material at [URL] for a
  detailed discussion of the Monte Carlo simulation and for additional
  experimental results.}\BibitemShut {Stop}%
\bibitem [{\citenamefont {Fenton}\ and\ \citenamefont
  {Warburton}(2008)}]{Fenton2008}%
  \BibitemOpen
  \bibfield  {author} {\bibinfo {author} {\bibfnamefont {J.~C.}\ \bibnamefont
  {Fenton}}\ and\ \bibinfo {author} {\bibfnamefont {P.~A.}\ \bibnamefont
  {Warburton}},\ }\href {https://doi.org/10.1103/PhysRevB.78.054526} {\bibfield
   {journal} {\bibinfo  {journal} {Phys. Rev. B}\ }\textbf {\bibinfo {volume}
  {78}},\ \bibinfo {pages} {054526} (\bibinfo {year} {2008})}\BibitemShut
  {NoStop}%
\bibitem [{\citenamefont {Longobardi}\ \emph
  {et~al.}(2011{\natexlab{a}})\citenamefont {Longobardi}, \citenamefont
  {Massarotti}, \citenamefont {Rotoli}, \citenamefont {Stornaiuolo},
  \citenamefont {Papari}, \citenamefont {Kawakami}, \citenamefont {Piero~Pepe},
  \citenamefont {Barone},\ and\ \citenamefont {Tafuri}}]{Longobardi2011}%
  \BibitemOpen
  \bibfield  {author} {\bibinfo {author} {\bibfnamefont {L.}~\bibnamefont
  {Longobardi}}, \bibinfo {author} {\bibfnamefont {D.}~\bibnamefont
  {Massarotti}}, \bibinfo {author} {\bibfnamefont {G.}~\bibnamefont {Rotoli}},
  \bibinfo {author} {\bibfnamefont {D.}~\bibnamefont {Stornaiuolo}}, \bibinfo
  {author} {\bibfnamefont {G.}~\bibnamefont {Papari}}, \bibinfo {author}
  {\bibfnamefont {A.}~\bibnamefont {Kawakami}}, \bibinfo {author}
  {\bibfnamefont {G.}~\bibnamefont {Piero~Pepe}}, \bibinfo {author}
  {\bibfnamefont {A.}~\bibnamefont {Barone}},\ and\ \bibinfo {author}
  {\bibfnamefont {F.}~\bibnamefont {Tafuri}},\ }\href
  {https://doi.org/10.1063/1.3624471} {\bibfield  {journal} {\bibinfo
  {journal} {Appl. Phys. Lett.}\ }\textbf {\bibinfo {volume} {99}},\ \bibinfo
  {pages} {062510} (\bibinfo {year} {2011}{\natexlab{a}})}\BibitemShut
  {NoStop}%
\bibitem [{\citenamefont {Longobardi}\ \emph
  {et~al.}(2011{\natexlab{b}})\citenamefont {Longobardi}, \citenamefont
  {Massarotti}, \citenamefont {Rotoli}, \citenamefont {Stornaiuolo},
  \citenamefont {Papari}, \citenamefont {Kawakami}, \citenamefont {Pepe},
  \citenamefont {Barone},\ and\ \citenamefont {Tafuri}}]{Longobardi2011a}%
  \BibitemOpen
  \bibfield  {author} {\bibinfo {author} {\bibfnamefont {L.}~\bibnamefont
  {Longobardi}}, \bibinfo {author} {\bibfnamefont {D.}~\bibnamefont
  {Massarotti}}, \bibinfo {author} {\bibfnamefont {G.}~\bibnamefont {Rotoli}},
  \bibinfo {author} {\bibfnamefont {D.}~\bibnamefont {Stornaiuolo}}, \bibinfo
  {author} {\bibfnamefont {G.}~\bibnamefont {Papari}}, \bibinfo {author}
  {\bibfnamefont {A.}~\bibnamefont {Kawakami}}, \bibinfo {author}
  {\bibfnamefont {G.~P.}\ \bibnamefont {Pepe}}, \bibinfo {author}
  {\bibfnamefont {A.}~\bibnamefont {Barone}},\ and\ \bibinfo {author}
  {\bibfnamefont {F.}~\bibnamefont {Tafuri}},\ }\href
  {https://doi.org/10.1103/PhysRevB.84.184504} {\bibfield  {journal} {\bibinfo
  {journal} {Phys. Rev. B}\ }\textbf {\bibinfo {volume} {84}},\ \bibinfo
  {pages} {184504} (\bibinfo {year} {2011}{\natexlab{b}})}\BibitemShut
  {NoStop}%
\bibitem [{\citenamefont {Martinis}\ and\ \citenamefont
  {Kautz}(1989)}]{Martinis1989}%
  \BibitemOpen
  \bibfield  {author} {\bibinfo {author} {\bibfnamefont {J.~M.}\ \bibnamefont
  {Martinis}}\ and\ \bibinfo {author} {\bibfnamefont {R.~L.}\ \bibnamefont
  {Kautz}},\ }\href {https://doi.org/10.1103/PhysRevLett.63.1507} {\bibfield
  {journal} {\bibinfo  {journal} {Phys. Rev. Lett.}\ }\textbf {\bibinfo
  {volume} {63}},\ \bibinfo {pages} {1507} (\bibinfo {year}
  {1989})}\BibitemShut {NoStop}%
\bibitem [{\citenamefont {Iansiti}\ \emph {et~al.}(1987)\citenamefont
  {Iansiti}, \citenamefont {Johnson}, \citenamefont {Smith}, \citenamefont
  {Rogalla}, \citenamefont {Lobb},\ and\ \citenamefont
  {Tinkham}}]{Iansiti1987}%
  \BibitemOpen
  \bibfield  {author} {\bibinfo {author} {\bibfnamefont {M.}~\bibnamefont
  {Iansiti}}, \bibinfo {author} {\bibfnamefont {A.~T.}\ \bibnamefont
  {Johnson}}, \bibinfo {author} {\bibfnamefont {W.~F.}\ \bibnamefont {Smith}},
  \bibinfo {author} {\bibfnamefont {H.}~\bibnamefont {Rogalla}}, \bibinfo
  {author} {\bibfnamefont {C.~J.}\ \bibnamefont {Lobb}},\ and\ \bibinfo
  {author} {\bibfnamefont {M.}~\bibnamefont {Tinkham}},\ }\href
  {https://doi.org/10.1103/PhysRevLett.59.489} {\bibfield  {journal} {\bibinfo
  {journal} {Phys. Rev. Lett.}\ }\textbf {\bibinfo {volume} {59}},\ \bibinfo
  {pages} {489} (\bibinfo {year} {1987})}\BibitemShut {NoStop}%
\bibitem [{\citenamefont {Iansiti}\ \emph {et~al.}(1989)\citenamefont
  {Iansiti}, \citenamefont {Tinkham}, \citenamefont {Johnson}, \citenamefont
  {Smith},\ and\ \citenamefont {Lobb}}]{Iansiti1989}%
  \BibitemOpen
  \bibfield  {author} {\bibinfo {author} {\bibfnamefont {M.}~\bibnamefont
  {Iansiti}}, \bibinfo {author} {\bibfnamefont {M.}~\bibnamefont {Tinkham}},
  \bibinfo {author} {\bibfnamefont {A.~T.}\ \bibnamefont {Johnson}}, \bibinfo
  {author} {\bibfnamefont {W.~F.}\ \bibnamefont {Smith}},\ and\ \bibinfo
  {author} {\bibfnamefont {C.~J.}\ \bibnamefont {Lobb}},\ }\href
  {https://doi.org/10.1103/PhysRevB.39.6465} {\bibfield  {journal} {\bibinfo
  {journal} {Phys. Rev. B}\ }\textbf {\bibinfo {volume} {39}},\ \bibinfo
  {pages} {6465} (\bibinfo {year} {1989})}\BibitemShut {NoStop}%
\bibitem [{Note1()}]{Note1}%
  \BibitemOpen
  \bibinfo {note} {Each SPD was obtained by recording 5,000 switching
  events.}\BibitemShut {Stop}%
\bibitem [{\citenamefont {Beenakker}\ and\ \citenamefont {van
  Houten}(1991)}]{Beenakker1991}%
  \BibitemOpen
  \bibfield  {author} {\bibinfo {author} {\bibfnamefont {C.~W.~J.}\
  \bibnamefont {Beenakker}}\ and\ \bibinfo {author} {\bibfnamefont
  {H.}~\bibnamefont {van Houten}},\ }\href
  {https://doi.org/10.1103/PhysRevLett.66.3056} {\bibfield  {journal} {\bibinfo
   {journal} {Phys. Rev. Lett.}\ }\textbf {\bibinfo {volume} {66}},\ \bibinfo
  {pages} {3056} (\bibinfo {year} {1991})}\BibitemShut {NoStop}%
\bibitem [{\citenamefont {Massarotti}\ \emph {et~al.}(2012)\citenamefont
  {Massarotti}, \citenamefont {Longobardi}, \citenamefont {Galletti},
  \citenamefont {Stornaiuolo}, \citenamefont {Montemurro}, \citenamefont
  {Pepe}, \citenamefont {Rotoli}, \citenamefont {Barone},\ and\ \citenamefont
  {Tafuri}}]{Massarotti2012}%
  \BibitemOpen
  \bibfield  {author} {\bibinfo {author} {\bibfnamefont {D.}~\bibnamefont
  {Massarotti}}, \bibinfo {author} {\bibfnamefont {L.}~\bibnamefont
  {Longobardi}}, \bibinfo {author} {\bibfnamefont {L.}~\bibnamefont
  {Galletti}}, \bibinfo {author} {\bibfnamefont {D.}~\bibnamefont
  {Stornaiuolo}}, \bibinfo {author} {\bibfnamefont {D.}~\bibnamefont
  {Montemurro}}, \bibinfo {author} {\bibfnamefont {G.}~\bibnamefont {Pepe}},
  \bibinfo {author} {\bibfnamefont {G.}~\bibnamefont {Rotoli}}, \bibinfo
  {author} {\bibfnamefont {A.}~\bibnamefont {Barone}},\ and\ \bibinfo {author}
  {\bibfnamefont {F.}~\bibnamefont {Tafuri}},\ }\href
  {https://doi.org/10.1063/1.3699625} {\bibfield  {journal} {\bibinfo
  {journal} {Low Temperature Physics}\ }\textbf {\bibinfo {volume} {38}},\
  \bibinfo {pages} {263} (\bibinfo {year} {2012})},\ \Eprint
  {https://arxiv.org/abs/https://doi.org/10.1063/1.3699625}
  {https://doi.org/10.1063/1.3699625} \BibitemShut {NoStop}%
\bibitem [{\citenamefont {Paul}()}]{Paul2007}%
  \BibitemOpen
  \bibfield  {author} {\bibinfo {author} {\bibfnamefont {C.~R.}\ \bibnamefont
  {Paul}},\ }\href@noop {} {\emph {\bibinfo {title} {Analysis of Multiconductor
  Transmission Lines}}}\ (\bibinfo  {publisher} {Wiley-IEEE Press})\BibitemShut
  {NoStop}%
\bibitem [{\citenamefont {Devoret}\ \emph {et~al.}()\citenamefont {Devoret},
  \citenamefont {Martinis},\ and\ \citenamefont {Clarke}}]{Devoret1985}%
  \BibitemOpen
  \bibfield  {author} {\bibinfo {author} {\bibfnamefont {M.~H.}\ \bibnamefont
  {Devoret}}, \bibinfo {author} {\bibfnamefont {J.~M.}\ \bibnamefont
  {Martinis}},\ and\ \bibinfo {author} {\bibfnamefont {J.}~\bibnamefont
  {Clarke}},\ }\href {https://doi.org/10.1103/PhysRevLett.55.1908} {\ \textbf
  {\bibinfo {volume} {55}},\ \bibinfo {pages} {1908}}\BibitemShut {NoStop}%
\bibitem [{\citenamefont {Kautz}\ and\ \citenamefont {Martinis}()}]{Kautz1990}%
  \BibitemOpen
  \bibfield  {author} {\bibinfo {author} {\bibfnamefont {R.~L.}\ \bibnamefont
  {Kautz}}\ and\ \bibinfo {author} {\bibfnamefont {J.~M.}\ \bibnamefont
  {Martinis}},\ }\href {https://doi.org/10.1103/PhysRevB.42.9903} {\ \textbf
  {\bibinfo {volume} {42}},\ \bibinfo {pages} {9903}}\BibitemShut {NoStop}%
\bibitem [{\citenamefont {Gloos}\ and\ \citenamefont {Anders}()}]{Gloos1999}%
  \BibitemOpen
  \bibfield  {author} {\bibinfo {author} {\bibfnamefont {K.}~\bibnamefont
  {Gloos}}\ and\ \bibinfo {author} {\bibfnamefont {F.}~\bibnamefont {Anders}},\
  }\href {https://doi.org/10.1023/A:1021874709355} {\ \textbf {\bibinfo
  {volume} {116}},\ \bibinfo {pages} {21}}\BibitemShut {NoStop}%
\bibitem [{\citenamefont {Golubov}\ \emph {et~al.}()\citenamefont {Golubov},
  \citenamefont {Kupriyanov},\ and\ \citenamefont {Il'ichev}}]{Golubov2004}%
  \BibitemOpen
  \bibfield  {author} {\bibinfo {author} {\bibfnamefont {A.~A.}\ \bibnamefont
  {Golubov}}, \bibinfo {author} {\bibfnamefont {M.~Y.}\ \bibnamefont
  {Kupriyanov}},\ and\ \bibinfo {author} {\bibfnamefont {E.}~\bibnamefont
  {Il'ichev}},\ }\href {https://doi.org/10.1103/RevModPhys.76.411} {\ \textbf
  {\bibinfo {volume} {76}},\ \bibinfo {pages} {411}}\BibitemShut {NoStop}%
\end{thebibliography}%

\clearpage
\newpage
\newcounter{myc} 
\newcounter{myc2} 
\renewcommand{\thefigure}{S.\arabic{myc}}
\renewcommand{\theequation}{S.\arabic{myc2}}

\section{Supplemental Material}

\subsection{Sample Fabrication}
Samples were fabricated from a heterostructure grown on InP (001) substrates by molecular beam epitaxy techniques. The heterostructure consists of a step-graded metamorphic InAlAs buffer and a $7~\mathrm{nm}$ thick InAs quantum well, confined by $\mathrm{In_{0.75}Ga_{0.25}As}$ barriers $10~\mathrm{nm}$ below the surface. A $10~\mathrm{nm}$ thick Al layer is deposited on top of the heterostructure, in the same chamber as the III-V growth while maintaining vacuum. The peak mobility is $18000~\mathrm{cm^{2}V^{-1}s^{-1}}$ for an electron density of $n=8\cdot10^{11}~\mathrm{cm^{-2}}$. This gives an electron mean free path of $l_{e}\gtrapprox350~\mathrm{nm}$, hence we expect all JJs measured here to be ballistic along the length $L$ of the junction.

The sample is defined by first isolating the large mesa structures on which the device is patterned. This is done by selectively removing the top Al layer (with Al etchant Transene D) before etching $\sim250~\mathrm{nm}$ into the III-V heterostructure using a chemical wet etch ($220:55:3:3$ solution of $\mathrm{H_{2}O:C_{6}H_{8}O_{7}:H_{3}PO_{4}:H_{2}O_{2}}$). The Al device is then patterned on top of the mesa, by selective etching of the Al with Transene D at $50~\mathrm{^{\circ}C}$ for $4~\mathrm{s}$. To control the exposed III-V region, we deposit a $15~\mathrm{nm}$ layer of $\mathrm{HfO_{2}}$ by atomic layer deposition, before evaporating metallic gate electrodes. These are deposited in two steps: the first consists of $5~\mathrm{nm}$ Ti and $20~\mathrm{nm}$ of Au on top of the device region; the second, $10~\mathrm{nm}$ of Ti and $350~\mathrm{nm}$ of Al, contacts the gates on top of the mesa to the bonding pads. 

\subsection{Lock-in Measurements}
\setcounter{myc}{1}
\begin{figure}
	\includegraphics[width=\columnwidth]{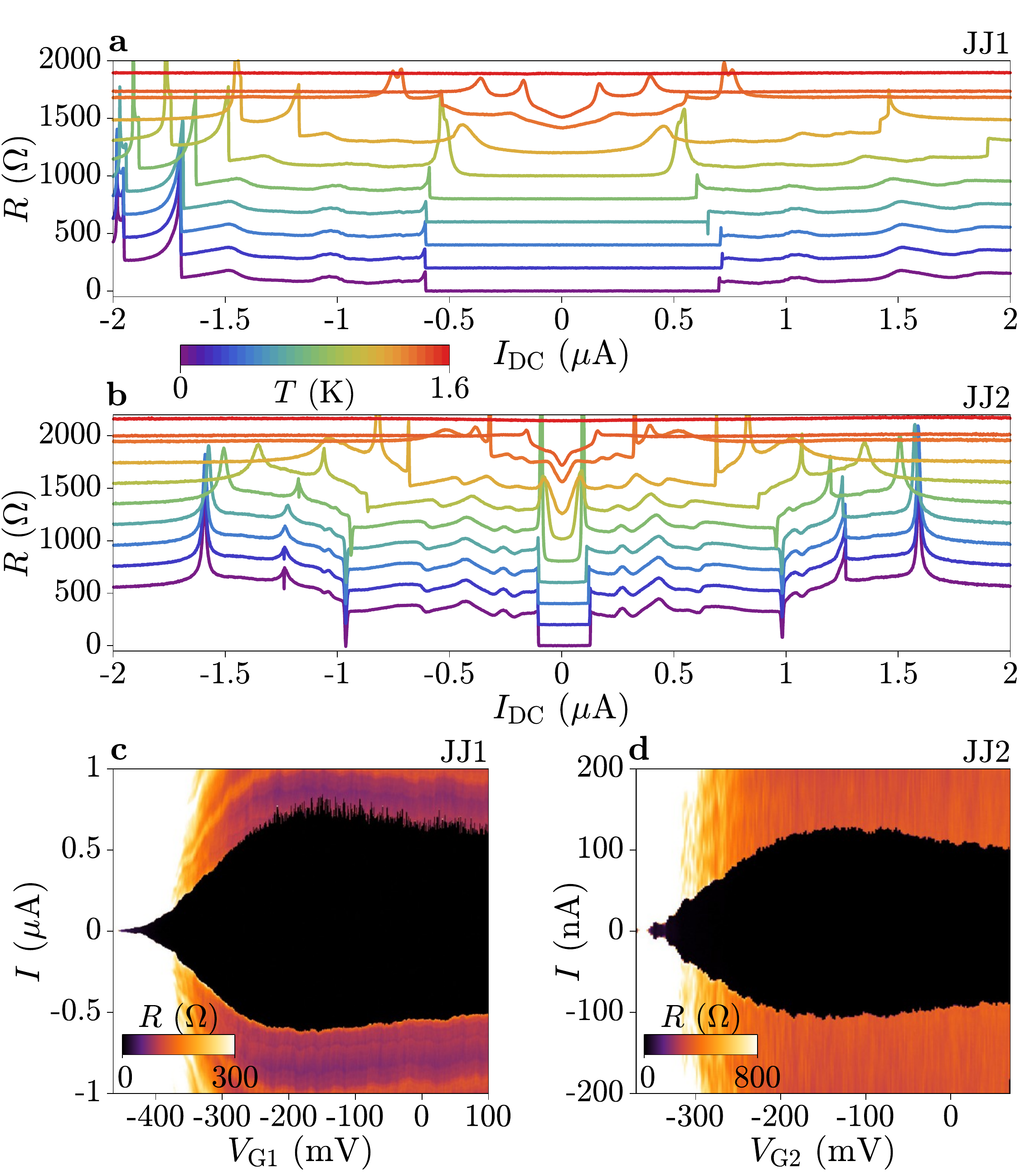}
	\caption{(a) Current-biased measurement of differential resistance $R$ of JJ1 with an applied bias of $\Idc$, in the positive sweep direction. (b) Same as (a) for JJ2. (c) Differential resistance $R$ of JJ1 as a function of gate voltage $\Vgone$. The junction goes from fully open to fully closed in the gate voltage range. (d) Same as (c) for JJ2. Note that the color scale in (c) and (d) is saturated to highlight the normal state resistance $R_{N}$ close to zero gate voltage.}
	\label{Sfig1}
\end{figure}

Electrical measurements were performed in a dilution refrigerator with a mixing chamber temperature below $20~\mathrm{mK}$. Initial characterization was performed by standard lock-in techniques. An AC current of $I_{\mathrm{AC}}=2~\mathrm{nA}$ was applied to the source contact of the SQUID device, with a frequency of $233~\mathrm{Hz}$. The four-terminal differential voltage $V_{\mathrm{AC}}$ across the SQUID was measured at this frequency, via a differential voltage amplifier with $1000$ times gain. The differential resistance $R=V_{\mathrm{AC}}/I_{\mathrm{AC}}$ was measured as a function of the applied DC current $\Idc$. 

In addition to the $\Bperp$-dependent measurements presented in Fig.~1 of the Main Text, we show temperature- and gate-dependent measurements of JJ1 and JJ2. In Figs.~\ref{Sfig1}(a) and (b), for JJ1 and JJ2 respectively, we show the differential resistance $R$ as a function of $\Idc$, swept from negative to positive currents. The color denotes the temperature, which ranges from $20~\mathrm{mK}$ to $1.6~\mathrm{K}$, at which point both JJs are fully resistive. We offset the vertical axis by $200~\mathrm{\Omega}$ between each temperature trace, to highlight the zero-resistance state at low bias currents.

As we increase $\Idc$ from $-2~\mathrm{\mu A}$, each junction undergoes a transition from the resistive to superconducting state at the retrapping current $I_{\mathrm{R}}$. At positive bias, the superconducting-to-resistive transition occurs at the switching current $\Isw$. The difference between the two, most notable in JJ1, indicates the underdamped JJ behavior. At high temperatures the superconducting state softens, leading to a finite resistance at bias values below $\Isw$. This is expected from phase diffusive JJs at high temperatures, but makes determination of $\Isw$ less reliable; hence we do not present SPDs at temperatures $T>1~\mathrm{K}$, where this effect is significant.

Figures~\ref{Sfig1}(c) and (d) show the gate dependence of the differential resistance $R$ across JJ1 and JJ2, respectively. The normal state resistances for JJ1 and JJ2 are $R_{N,1}=150~\mathrm{\Omega}$ and $R_{N,2}=540~\mathrm{\Omega}$, respectively. At a small negative gate voltage, the switching current reaches its maximum. The peak occurs at $\Vgone=-180~\mathrm{mV}$ for JJ1 and $\Vgtwo=-140~\mathrm{mV}$ for JJ2. These define the operating points for each junction in Figs.~1-3 of the Main Text. Each JJ can be tuned to the completely resistive state with sufficiently negative gate voltages, as seen in Figs.~\ref{Sfig1}(c) and (d) for JJ1 and JJ2 respectively. To measure a single junction in isolation, we apply $V_{G}<-400~\mathrm{mV}$ to the other junction so that no supercurrent flows there, as in Figs.~1,2 of the Main Text. In Fig.~\ref{Sfig1}(c), we observe large fluctuations in the switching current for $\Vgone>-200~\mathrm{mV}$. This is indicative of large quantum fluctuations when the critical current is large.

\subsection{Phase Escape in Single JJs}
Phase dynamics in Josephson junctions are often described using the analogy of a massive phase particle in a tilted washboard potential, where the particle mass corresponds to the JJ  capacitance and its damping to the inverse of its resistance. The zero-resistance state corresponds to the particle being trapped in an energy minimum while the resistive state corresponds to the particle moving along the potential. A transition to the resistive state generally takes place, via quantum or thermal fluctuations, at switching currents $\Isw$ lower than the JJ critical current $\Ic$. To measure this transition, we apply a sawtooth signal from a waveform generator through a bias resistor to rapidly ramp $\Idc$, and detect a switch to the resistive state by measuring the point at which the voltage across the JJ increases above a threshold. The value of $\Idc$ at which this switch occurs is recorded as the switching current, $\Isw$. We measure $\Isw$ more than 10,000 times and collect the results in a switching probability distribution (SPD). The SPDs are recorded as a function of temperature, from $20~\mathrm{mK}$ to $1~\mathrm{K}$.

\setcounter{myc}{2}
\begin{figure}
	\includegraphics[width=\columnwidth]{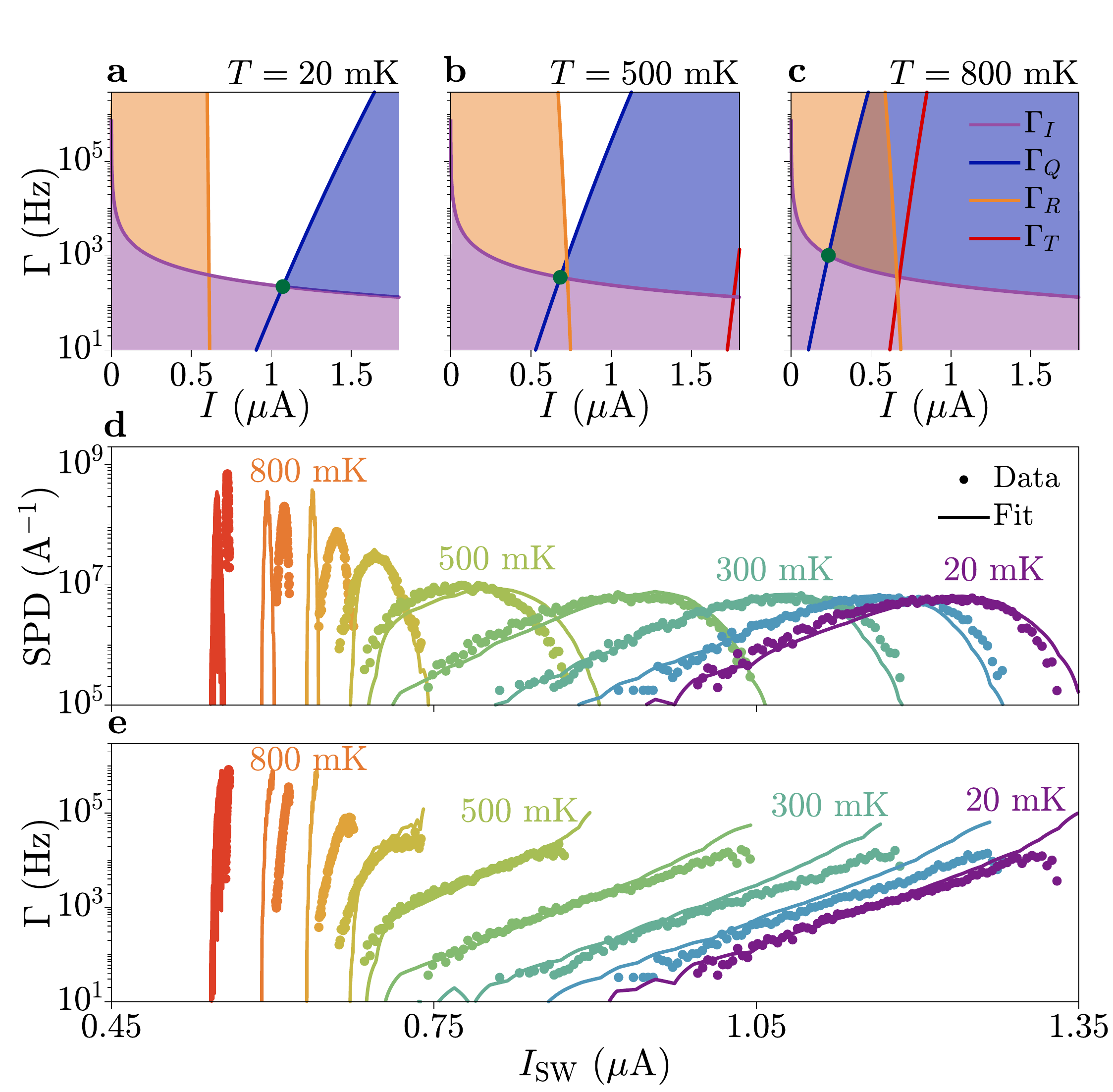}
	\caption{(a-c) Simulated escape rates for $T=20~\mathrm{mK}$, $500~\mathrm{mK}$ and $800~\mathrm{mK}$ respectively, for the fit parameters of JJ1. Bias-current ramp-rate $\GI$, MQT escape rate $\GQ$, retrapping rate $\GR$ and thermal escape rate $\GT$ are compared across the bias range $I$. The green dot indicates $\GI=\GQ$, at which point escape events are measurable. (d) SPDs of JJ1: Experimental data (points) compared with Monte Carlo simulation (lines) for temperatures $20~\mathrm{mK}$ to $1~\mathrm{K}$. The temperature is indicated by the color, as in the Main Text. (e) Escape rates corresponding to the SPDs in (d).}
	\label{Sfig2}
\end{figure}

The SPDs are dependent on the rate at which the DC current is increased, $\nu$. To compare with theoretical expressions for phase dynamic mechanisms, we must convert to a measurement-independent quantity: the escape rate, $\Gamma$. This is done via KFD transform using the following equation \cite{Bezryadin2012}:
\setcounter{myc2}{1}
\begin{equation}
	\Gamma(\Isw) = \mathrm{SPD}(\Isw) \nu \left[1- \int_{0}^{\Isw} \mathrm{SPD}(I)dI \right]^{-1}
	\label{eq:S1}
\end{equation} 
where $\nu$ is the ramp rate of the DC current, which was $\nu=240~\mathrm{\mu As^{-1}}$ for all measurements. 

In the following, we will consider equations derived for JJs with a sinusoidal current-phase relation. While our JJs are known to contain highly-transmissive modes, leading to deviations from this sinusoidal behavior, we use the existing theory in the absence of an alternative model and expect only small numerical deviation from the results presented.

The rate of escape of a Josephson junction to the resistive state under the action of a DC bias follows the general dependence \cite{Massarotti2012}:
\setcounter{myc2}{2}
\begin{equation}
	\Gamma(\Isw,T)=\Omega(\Isw,T)e^{-\dU(\Isw,T)/k_{B}T} ,
	\label{eq:2}
\end{equation}
where $\Omega$ is the attempt frequency and $\Delta U$ is the potential barrier height. Under a reduced DC bias of $\gamma=\Idc/\Ic$, the approximate barrier height is $\dU = 2\Ej\left(\sqrt{1-\gamma^{2}}-\gamma\cos^{-1}\gamma\right)$, with a Josephson energy of $\Ej=\hbar \Ic/2e$.

A transition to the resistive state can be promoted, for currents lower than $\Ic$, by macroscopic quantum tunnelling (MQT) or thermal activation (TA) across the barrier $\dU$. In the case of MQT, equation \ref{eq:S1} is adjusted to the analytical formula for escape \cite{Massarotti2012}
\setcounter{myc2}{3}
\begin{equation}
	\GQ=\frac{\plasma}{2\pi}\sqrt{\frac{864\pi\dU}{\hbar\plasma}}\mathrm{exp}\left( -7.2\left(1+\frac{0.87}{Q}\right)\frac{\dU}{\hbar \plasma}\right),
	\label{eq:3}
\end{equation}
where $\plasma=\omega_{P0}\left(1-\gamma^{2}\right)^{1/4}$ is the plasma frequency at $\gamma$ and $Q$ is the quality factor of the junction. The junction capacitance $C$ enters in the bare plasma frequency, $\omega_{P0} = \sqrt{2e\Ic/\hbar C}$. 

When dissipation is large it is possible for the junction to transition from the resistive to the superconducting state, referred to as retrapping. In this regime, many escape events are required to turn the junction resistive. This is referred to as phase diffusion (PD). For $Q\gg1$, we can use an analytical formula for this retrapping rate \cite{Massarotti2012}:
\setcounter{myc2}{4}
\begin{equation}
	\GR = \frac{\Isw-\Ir}{\Ic}\omega_{P0}\sqrt{\frac{\Ej}{2\pi k_{B}T}}\mathrm{exp}\left(-\left(\frac{\Isw-\Ir}{\Ic}\right)^{2}\frac{\Ej Q^{2}}{2k_{B}T}\right)
	\label{eq:4}
\end{equation} 
where $\Ir$ is the retrapping current. We measure intermediately damped junctions with $Q\gtrsim1$, so this relation can only be considered as an approximation.

The interplay between $\GQ$ and $\GR$ determines the phase escape regime: $\GQ\gg\GR$ in the MQT regime, whereas the reverse is true for phase diffusion. The rates depend on the junction properties: the critical current $\Ic$, the capacitance $C$ and the quality factor, which has the zero-temperature value $Q_{0}$. We show in Figs.~\ref{Sfig2}(a-c) the relevant rates in the system for the fit parameters for JJ1: $\Ic=3~\mathrm{\mu A}$, $C=1~\mathrm{fF}$ and $Q_{0}=7$, as discussed in more detail later. Each panel corresponds to a different temperature: $T=20~\mathrm{mK}$, $T=500~\mathrm{mK}$ and $T=800~\mathrm{mK}$, respectively. These highlight the change in the dominant regime from MQT to PD on increasing $T$. 

The ramp rate $\GI(I)=\nu/I$ defines the lowest frequencies at which an escape event can be measured, for a given bias current $I$. The point at which this intersects the escape rate gives the lowest bias current $I_{\mathrm{IE}}$ at which an escape event is measurable. Since quantum tunneling is the dominant escape mechanism, we define $ I_{\mathrm{IE}}$ as the current at which $\GI=\GQ$, marked in Figs.~\ref{Sfig2}(a-c) with the green dot. The retrapping rate $\GR$ is large for low bias currents but quickly decreases with an increase in $I$. Retrapping is significant when $\GR$ reaches a similar magnitude as $\GQ$. Escape by thermal activation is given by $\GT$, and is much smaller than $\GQ$ for all temperatures.

At the lowest temperature, escape by quantum tunneling dominates. This is clear since $\GQ\gg\GR$ for $I>I_{\mathrm{IE}}$. No retrapping of the phase occurs: a single escape event is sufficient to transition to the resistive state. At $T=500~\mathrm{mK}$, $\GQ\lesssim\GR$ close to $I_{\mathrm{IE}}$. For these low bias currents, the retrapping probability is high so the probability of escape in the junction is reduced relative to quantum tunneling alone. However, $\GQ\gg\GR$ at larger bias so escape occurs unhindered by phase diffusion. As the temperature increases, the effect of retrapping becomes more significant. At the high temperature of $800~\mathrm{mK}$, phase diffusion is dominant since $\GQ\ll\GR$ across the range of escape currents. 

The transition temperature $\Tstar$ between MQT and PD regimes is experimentally defined as the inflection point of the standard deviation as a function of temperature. This is the point at which the retrapping rate $\GR$ becomes dominant above MQT escape $\GQ$. We can specify this further as the temperature at which the escape, retrapping and ramp rates are equal, $T^{*}_{\mathrm{IER}}$. This is similar to Ref.~\cite{Fenton2008}. The trend of $\Imean$ with temperature, for the single junctions and the SQUID, changes at $\Tstar$ (see Figs.~2(e) and 3(c) in the Main Text). This suggests that our definition of $\Tstar$ is appropriate as a measure of the transition between MQT and PD regimes. 

The rates in Figs.~\ref{Sfig2}(a-c) are used to model the junction behavior in a Monte Carlo simulation, as in Ref.~\cite{Fenton2008}. As the DC bias current is increased, the simulated junction stochastically switches between the superconducting (0) and resistive (1) states according to the relative escape and retrapping rates. The junction is said to be resistive when the state, averaged over a window of current, exceeds 0.1. This process is performed 20,000 times and the generated $\Isw$ are combined into a switching probability distribution (SPD). 

The simulation uses the critical current $\Ic$, capacitance $C$ and quality factor $Q_{0}$ as input parameters for the rates $\GQ$ and $\GR$. We fit by hand the SPD at low temperature, obtaining $\Ic=3~\mathrm{\mu A}$, $C=1~\mathrm{fF}$ and $Q_{0}=7$. We use the Bardeen formula for the temperature dependence of $\Ic(T)  = \Ic(1-T^{2}/T_{C}^{2})^{3/2}$, with $T_{C}=1.18~\mathrm{K}$ from experimental results. Since $Q\propto \Ic^{1/2}$, we use $Q(T)=Q_{0}(1-T^{2}/T_{C}^{2})^{3/4}$.  We use the low-temperature fit result and the assumed temperature dependence to simulate the full dataset. Figures~\ref{Sfig2}(d) and (e) show the SPDs and escape rates for JJ1, respectively. The experimental data (circles) is fitted well by the simulated curve (lines). Deviation at high temperatures between the fit and the data is explained by the simple model used for the temperature dependence. Despite this, we capture the characteristic trend in the data.

\begin{figure}
	\includegraphics[width=\columnwidth]{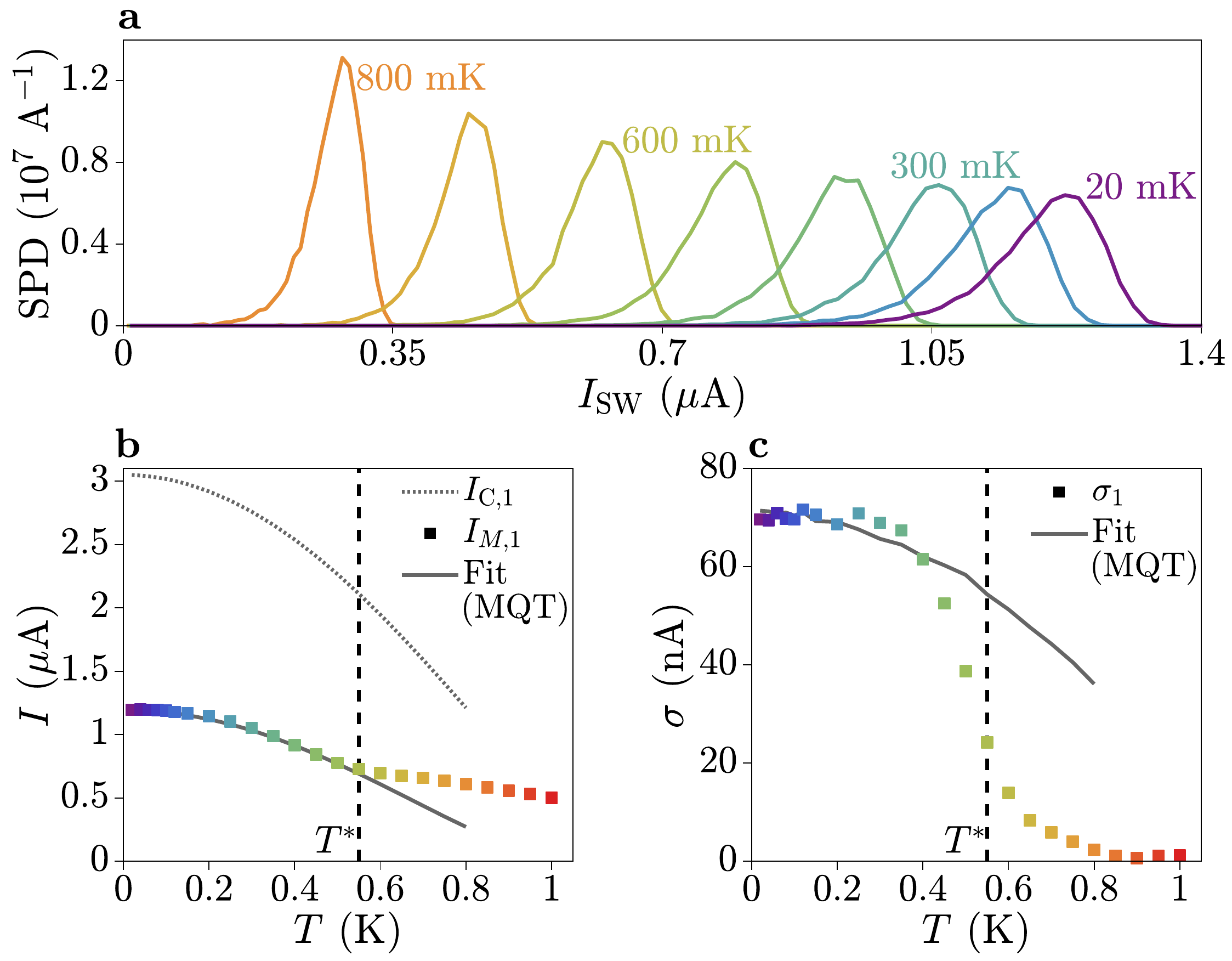}
	\caption{(a) Switching probability distributions (SPDs) obtained from a Monte Carlo simulation considering only escape by quantum tunneling (MQT), as a function of temperature. Calculated using the parameters $\Ic=3~\mathrm{\mu A}$, $C=1~\mathrm{fF}$ and $Q_{0}=7$. (b) Temperature dependence of critical current $\Icone$ (dashed line), JJ1 switching current $\Imone$ (squares) and MQT-only fit (solid line). (c) Temperature dependence of standard deviation in JJ1 (squares) and MQT-only fit (line).}
	\label{Sfig9}
\end{figure}

We now consider the impact of temperature on the switching current in the MQT regime. In Fig.~\ref{Sfig9} we calculate the SPD as a function of temperature using the same fit parameters as Fig.~\ref{Sfig2}, but considering only MQT escape without the presence of PD ($\GR=0$). This is instructive to see the impact of the temperature dependence of $\Ic$ and $Q$ on SPDs at low $T$. In particular, we see that the mean switching current $\Imone$ decreases with increasing temperature [Fig.~\ref{Sfig9}(a)]. This is due to the low $T_{C}$ of Al, which causes a change in $\Ic$ even at the lowest temperature of our experiment. This is evident in Fig.~\ref{Sfig9}(b) (dashed line). The experimental data for $\Imone$ (squares) aligns with the simulated result for MQT in the absence of PD (solid line) up to $T\sim\Tstar$. This is indicative of the dominance of MQT over PD up to $T\sim\Tstar$ in the experimental data. In Fig.~\ref{Sfig9}(c) we show the experimental standard deviation $\sone$ (squares) and the modelled standard deviation in the case of pure MQT (line). The slight temperature dependence of the fit result is due to the temperature dependence of $Q$, again a consequence of the low $T_{C}$ in Al.

\setcounter{myc}{3}

\setcounter{myc}{4}
\begin{figure}
	\includegraphics[width=\columnwidth]{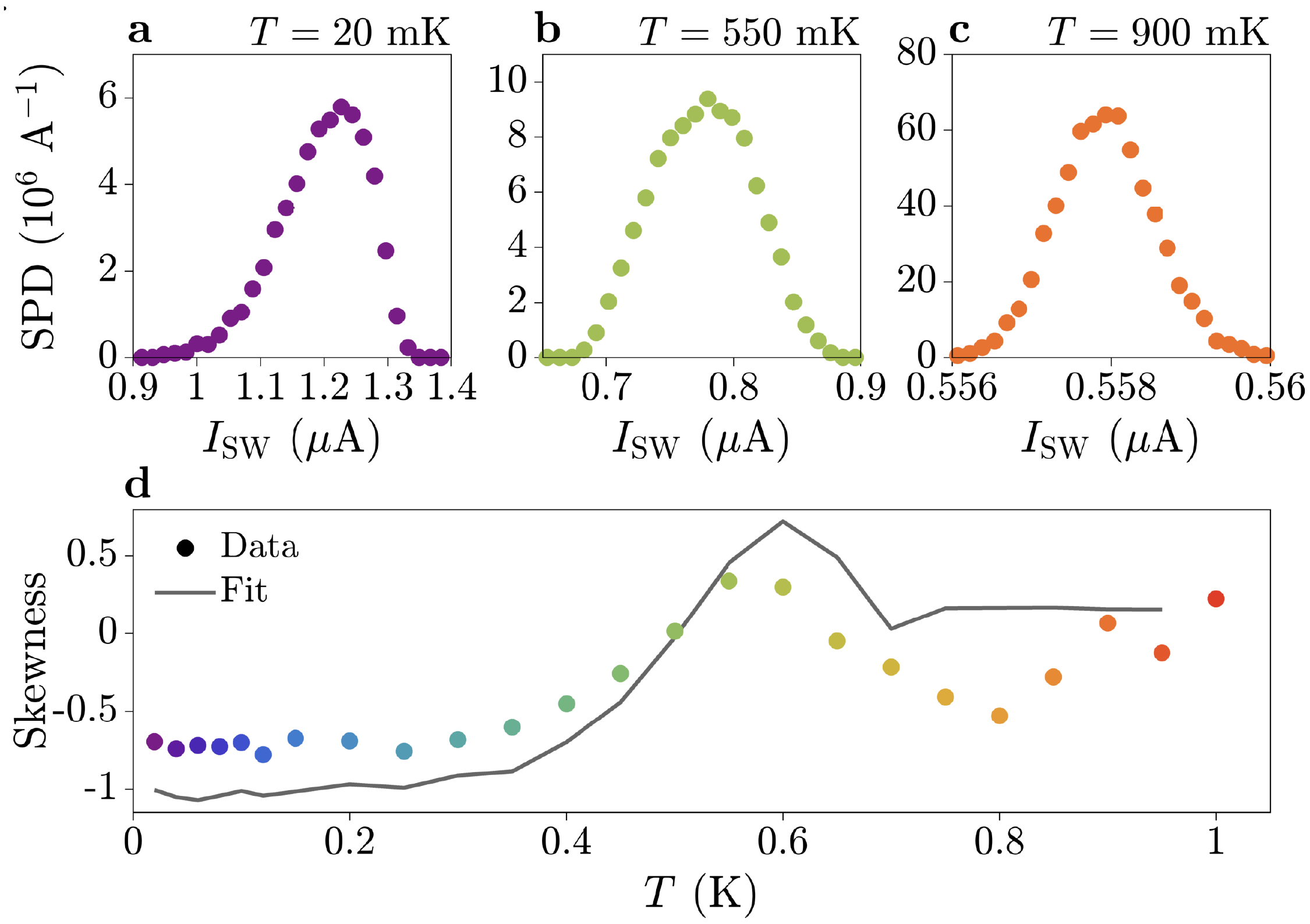}
	\caption{(a) Switching probability distribution (SPD) of JJ1 at $T=20~\mathrm{mK}$, plotted on a linear scale. Large negative skewness, consistent with a lack of phase diffusion at low temperature. (b) SPD at $T=550~\mathrm{mK}$, where skewness is positive. (c) Symmetric SPD at $T=900~\mathrm{mK}$. (d) Skewness of SPDs (circles) as a function of temperature, as indicated by the color. This is compared with the result from the Monte Carlo fit (gray curve).}
	\label{Sfig8}
\end{figure}

Further to the standard deviation, as shown in the Main Text, the skewness of SPDs is indicative of the phase escape mechanism. Phase escape unhindered by retrapping leads to a negative skewness close to -1~\cite{Murphy2013}. On entering the PD regime, the SPDs become more symmetric and the skewness tends towards zero. This is shown in Figs.~\ref{Sfig8}(a-c), where SPDs of JJ1 are plotted on a linear scale for $T=20~\mathrm{mK}$, $550~\mathrm{mK}$ and $900~\mathrm{mK}$ respectively. The trend in skewness is consistent with the interpretation above. At low temperature the SPD has large negative skewness, indicating that no phase diffusion is present. At higher temperatures, the SPDs are more symmetric with a skewness of zero at $T=900~\mathrm{mK}$. The skewness of JJ1 as a function of temperature (circles) is plotted in Fig.~\ref{Sfig8}(d), alongside the Monte Carlo fit result (gray line). Both the data and the Monte Carlo fit are negative at low temperature, and increase towards zero for higher temperatures where phase diffusion becomes dominant. 

In general, the quality factor of the junction is described by $Q=RC\omega_{P}$. For $\Ic=3~\mathrm{\mu A}$ and $C=1~\mathrm{fF}$ we get a resistance of $R=2.33~\mathrm{k\Omega}$, much larger than the normal state resistance at low frequency of $R_{N,1}=150~\mathrm{\Omega}$. We therefore conclude that damping at high frequency is relevant in the case of these junctions. For $Q\gg1$, we can relate the quality factor to the ratio of critical and retrapping currents: $Q=4\Ic/\pi\Ir$. For JJ1, $\Ir=600~\mathrm{nA}$ giving $Q=6.4$, close to the fit value.

Escape by quantum tunneling is dominant up to the critical temperature. By comparing Eqs.~\ref{eq:2} and~\ref{eq:3}, we obtain an effective temperature of quantum tunneling escape, 
\setcounter{myc2}{5}
\begin{equation}
	k_{B}T_{Q}=\frac{\hbar\omega_{P}}{7.2(1+0.87/Q)}.
	\label{eq:6}
\end{equation}
At low temperature, $T_{Q}\approx3~\mathrm{K}$ for JJ1. This exemplifies the large scale of quantum fluctuations relative to thermal excitations. The temperature at which $T_{Q}<T$ for the parameters of JJ1 is approximately $1~\mathrm{K}$: thermal activation is not significant up to the critical temperature.  

\subsection{Estimation of the Geometrical Capacitance}
The capacitance $C=1~\mathrm{fF}$ obtained in the Monte Carlo simulation is consistent with the geometrical capacitance of JJ1, calculated as the coplanar capacitance between epitaxial Al electrodes and given by the formula \cite{Paul2007}:
\setcounter{myc2}{6}
\begin{equation}
	C = \frac{\epsilon_{0}\epsilon_{r}W}{\pi}
	\ln\left[-2\left(\frac{\beta+1}{\beta-1}\right)\right],
	\label{eq:capacitor}
\end{equation}
where $\beta = \sqrt[4]{1-\frac{L^{2}}{(L+2L_{S})^{2}}}$. We use the dielectric constant $\epsilon_{r}=12.3$ of InAs at high frequency. The geometrical parameters of the junction are: lateral extent of JJ electrodes, $W=5~\mathrm{\mu m}$; separation of JJ electrodes, $L=40~\mathrm{nm}$; and length of superconducting leads, $L_{S}=250~\mathrm{nm}$. This gives $C=1.4~\mathrm{fF}$. The calculation does not consider the effect of the top gate electrode, which is grounded via a low-impedance terminal. It therefore acts as a screening layer for the electric field between the two junction electrodes. This reduces the capacitance, up to a factor of 2 for complete screening of half of the field between the Al electrodes. The presence of JJ2 in the SQUID loop contributes an additional shunt capacitance of $C_{2}=0.2~\mathrm{fF}$, for junction parameters $W_{2}=1.6~\mathrm{\mu m}$, $L_{2}=100~\mathrm{nm}$ and $L_{S,2}=250~\mathrm{nm}$. Screening of the field by the electrostatic gates may reduce this by up to a factor of 2. The contribution of JJ2 to the capacitance of JJ1 is therefore small, and does not modify the conclusions of the Main Text.

The large leads of the device might contribute as well to the shunt capacitance. Their geometrical capacitance, calculated with Eq.~\ref{eq:capacitor}, is about $3~\mathrm{fF}$. This value is likely decreased by the presence of the large gate electrodes between them. This analysis does not consider shunt capacitances present in the bonding pads and in the wiring of the dilution refrigerator. The extent to which shunt capacitances contributes to the plasma frequency is an open question, investigated in Refs.~\cite{Devoret1985,Kautz1990,Gloos1999}. The interpretation in the Main Text of a planar Josephson junction with a small capacitance $C$ is supported by the result of an analytical fit to Eq.~\ref{eq:3}, a Monte Carlo simulation of the temperature dependence, a consideration of the geometrical capacitance of the junction and a comparison to values previously reported in literature~\cite{Lee2011,Mayer2020}.

\subsection{Full SQUID Data and Monte Carlo Simulation}
\begin{figure}
 	\includegraphics[width=\columnwidth]{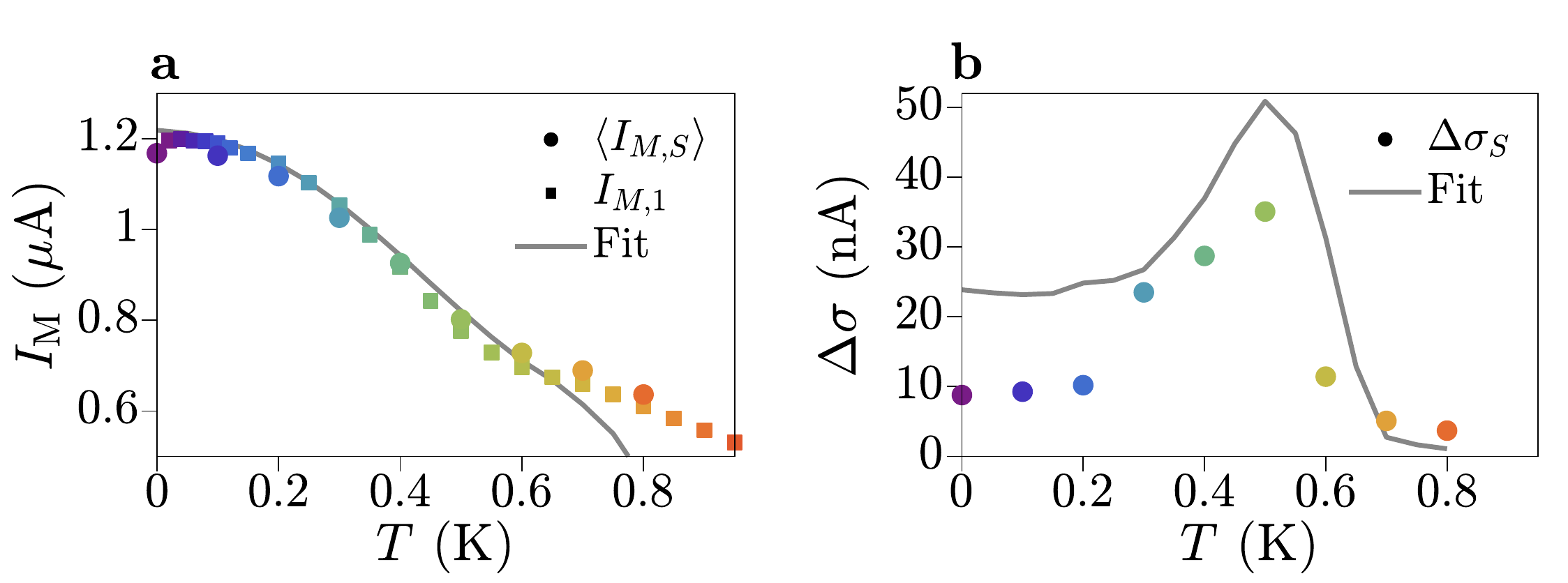}
 	\caption{(a) Average value of $\ImS$ across $\Bperp$ (circles), as a function of temperature (as indicated by the color), compared with the mean switching current of JJ1 in isolation $\Imone$ (squares) and a fit obtained by Monte Carlo simulation (gray curve). (b) Difference between the maximal and minimal $\sS$ across $\Bperp$, $\Delta\sS$ (circles), which indicates the size of the field-dependence in $\sS$ at a given temperature. This is compared with a Monte Carlo simulation (gray line), which reproduces the trend.}
 	\label{Sfig3}
 \end{figure}

Figure~3 in the Main Text shows the switching probability in the SQUID configuration, as a function of $\Bperp$ and $T$. From $\ImS$ and $\sS$ as a function of $\Bperp$ we extract $\Delta\ImS/2$ and $\langle\sS\rangle$. In Fig.~\ref{Sfig3}, we plot the remaining extracted parameters $\langle\ImS\rangle$ and $\Delta\sS$ (circles) with the corresponding results for the Monte Carlo simulation (lines). 

 Figure~\ref{Sfig3}(a) shows the field-averaged value of the oscillations in $\ImS$ (circles) compared with $\Imone$, the mean switching current for JJ1 (squares). The two align across all temperatures, confirming that JJ1 dominates the average SQUID behavior. 
 
 \setcounter{myc}{5}

 The Monte Carlo simulation (gray line) follows the data at low temperatures, but some deviations emerge above $T=600~\mathrm{mK}$ when the junction is almost completely phase diffusive. The escape rates are particularly sensitive to the damping $Q$ in this regime, so deviations between simulations and experiment at high temperature might be accounted for with a more complex temperature dependence. Figure~3(c) of the Main Text shows a similar deviation in $\Delta\ImS/2$, most likely from the same effect.

We observe a strong magnetic-field-dependence of the standard deviation in the SQUID, which is characterized by $\Delta\sigma_{S}$ in Fig.~\ref{Sfig3}(b). Close to the transition temperature $T^{*}\approx 0.55~\mathrm{K}$, $\Delta\sS$ is more than $35~\mathrm{nA}$. Both at low and high temperature, $\sS$ is almost constant across magnetic field. Oscillations in the simulated $\sS$ (gray lines) were observed at low-temperature, contrary to the experimental data, hence the larger simulated $\Delta\sS$.

The results are compared with the Monte Carlo simulation (lines). As described in the Main Text, the fit parameter $\IcS$ is obtained from the low-temperature data. Since the SQUID is in the MQT regime, we fit each escape rate with Eq.~\ref{eq:3} using a fixed $C=1~\mathrm{fF}$ and $Q_{0}=7$ from JJ1. The skewness of $\IcS$ indicates that highly transmissive modes are present in the junction. To capture this skewness, we consider a JJ with many modes of equal transmission $\bar{\tau}$:

\setcounter{myc2}{7}
\begin{equation}
	\IcS=\Icone+I_{0}\frac{\sin(\varphi)}{\sqrt{1-\bar{\tau}\sin^{2}(\varphi/2)}},
\end{equation}
with $\Icone=3~\mathrm{\mu A}$, $I_{0}=480~\mathrm{nA}$ and effective transmission $\bar{\tau}=0.77$. This gives $\Delta\IcS/2=650~\mathrm{nA}$. The mean value of $\IcS$ at $T=20~\mathrm{mK}$ is given by the solid line in Fig.~\ref{Sfig3}(a). We use the low-temperature result to simulate the full dataset by varying $\Icone$ and $I_{0}$ with a Bardeen dependence, as for the isolated junctions. 

We can further quantify the properties of JJ1 and JJ2 by considering their $I_{\mathrm{C}}R_{N}$ product~\cite{Golubov2004}. This gives $\Icone\cdot R_{N,1}=3~\mathrm{\mu A}\cdot150~\mathrm{\Omega} = 450~\mathrm{\mu V}$ and $\Ictwo\cdot R_{N,2}=650~\mathrm{nA}\cdot540~\mathrm{\Omega} = 350~\mathrm{\mu V}$, for JJ1 and JJ2 respectively. The expected $I_{\mathrm{C}}R_{N}$ product is $\pi\Delta/e$ in the clean limit, or $\pi\Delta/2e$ in the tunneling limit ($l_{e}<\xi$). For a superconducting gap of $\Delta=180~\mathrm{\mu eV}$, this gives $I_{\mathrm{C}}R_{N}=565~\mathrm{\mu V}$ or $283~\mathrm{\mu V}$, respectively. Hence, both JJ1 and JJ2 are in an intermediate regime between the clean and tunneling limits, with JJ2 closer to the tunneling limit than JJ1.

\subsection{$\Vgtwo$ Dependence}
\setcounter{myc}{6}
\begin{figure}
	\includegraphics[width=\columnwidth]{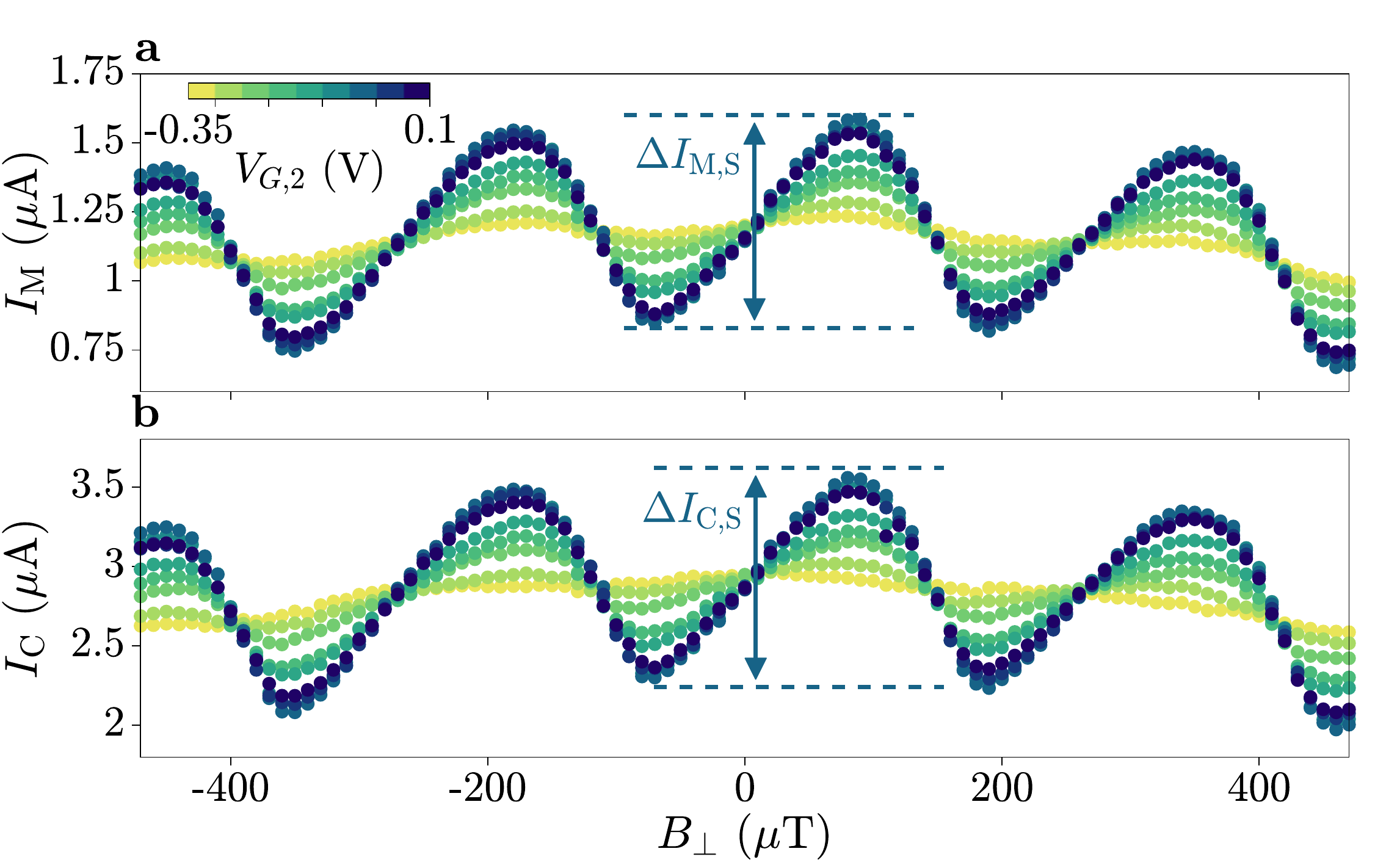}
	\caption{(a) Mean switching current $\ImS$ as a function of $\Bperp$, taken at $T=20~\mathrm{mK}$ for different values of the JJ2 gate voltage, $\Vgtwo$. The oscillation amplitude $\Delta\ImS/2$ gives the switching current of JJ2 in the SQUID, as indicated by the arrow for $\Vgtwo=-100~\mathrm{mV}$. (b) Critical current $\IcS$, obtained from the curves in (a) by fitting SPDs in the MQT regime. The oscillation amplitude $\Ictwo=\Delta\IcS/2$ gives the critical current of JJ2 as a function of $\Vgtwo$, as indicated by the arrow for $\Vgtwo=-100~\mathrm{mV}$}
	\label{Sfig4}
\end{figure}

In an asymmetric SQUID, the amplitude of oscillations as a function of $\Bperp$ is an indication of the current flowing through the small junction. This is an approximation to 
\setcounter{myc2}{7}
\begin{equation}
	\IcS=\sqrt{(\Icone-\Ictwo)^{2}+4\Icone\Ictwo\cos^{2}\left(\frac{\varphi-\varphi_{0}}{2}\right)},
\end{equation}
when $\Icone\gg\Ictwo$. We use this equation to extract the critical current of JJ2 in the asymmetric SQUID, as a function of gate voltage $\Vgtwo$.

We measure SQUID oscillations at $T=20~\mathrm{mK}$ and fixed $\Vgone=-180~\mathrm{mV}$, with different $\Vgtwo$ [see Fig.~\ref{Sfig4}(a)]. The SQUID is always in the MQT regime, independent of $\Vgtwo$, and the asymmetry is large. For each $\Vgtwo$, we extract $\Delta\ImS/2$ as shown, and plot the results in Fig.~4(a) of the Main Text. We extract the critical current $\IcS$ by fitting each SPD with Eq.~\ref{eq:3} in the MQT regime, with fixed capacitance $C$ and quality factor $Q_{0}$ defined by JJ1. The result is shown in Fig.~\ref{Sfig4}(b), from which we extract $\Ictwo=\Delta\IcS/2$.

\subsection{$\Vgone$ Dependence}
\begin{figure}
	\includegraphics[width=\columnwidth]{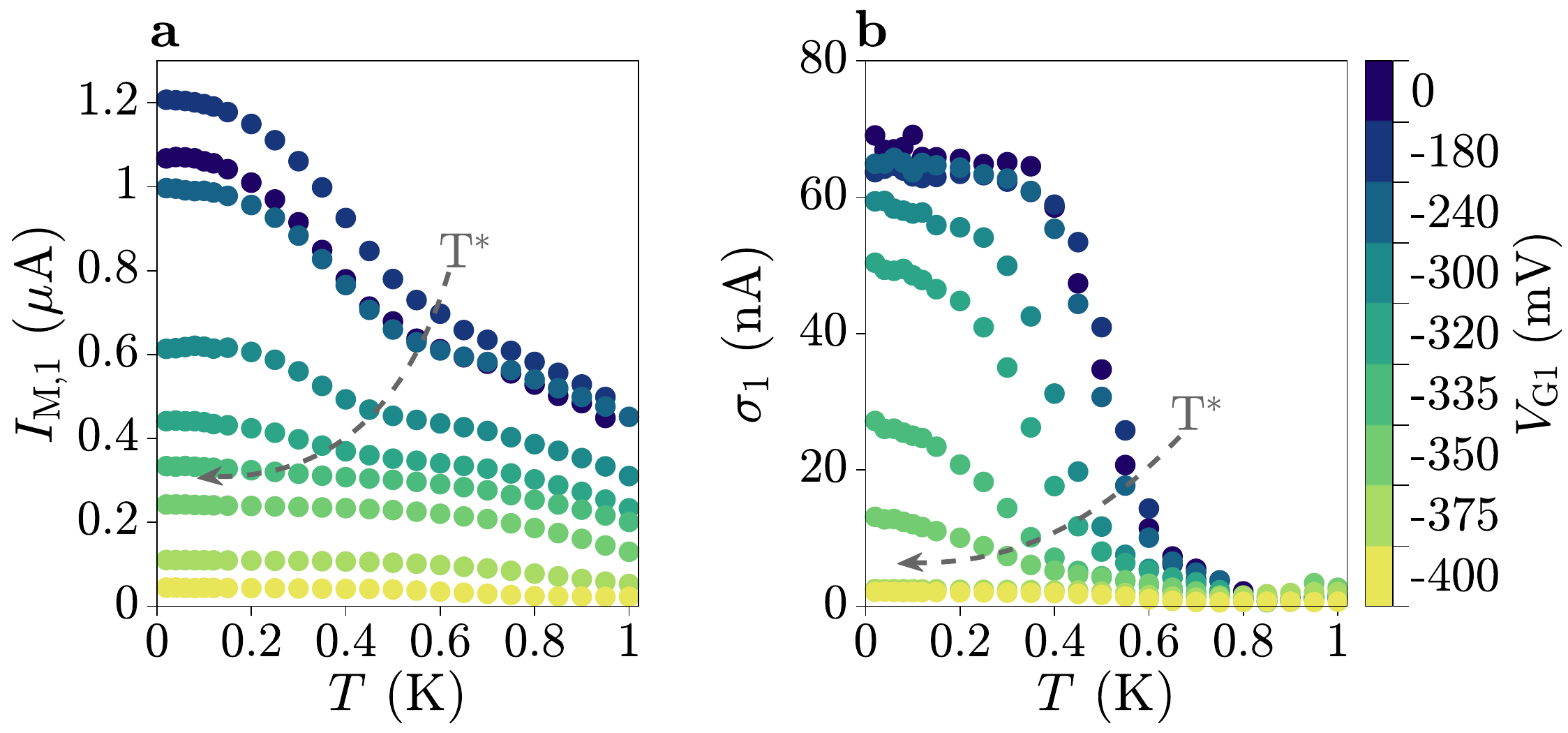}
	\caption{(a) Mean switching current $\Imone$ of switching probability distributions (SPDs) of JJ1 as a function of temperature. Measured for different gate voltages $\Vgone$, as indicated by the color. The decrease in transition temperature $\Tstar$ with $\Vgone$ is indicated by the arrow. (b) Standard deviation $\sone$ of SPDs of JJ1 as a function of temperature, for different gate voltages $\Vgone$. The decrease in $\Tstar$ with $\Vgone$ is indicated by the arrow.}
	\label{Sfig10}
\end{figure}

We tune the phase dynamics of JJ1 with a gate voltage $\Vgone$. We measure SPDs for JJ1 as a function of temperature at gate voltages spanning the range of critical currents. In Fig.~\ref{Sfig10}(a,b) we plot the mean and standard deviation of SPDs, respectively. As $\Vgone$ is tuned to more negative values, the transition temperature $\Tstar$ decreases until $\Vgone=-400~\mathrm{mV}$, where JJ1 is fully phase diffusive at the lowest temperature in the measurement. This transition is clear from both the mean switching current $\Imone$ and standard deviation $\sone$. At $\Vgone=-320~\mathrm{mV}$, $\sone$ at $T=20~\mathrm{mK}$ is $50~\mathrm{nA}$, reduced relative to $65~\mathrm{nA}$ at $\Vgone=-180~\mathrm{mV}$. Additionally, $\sone$ is not constant with temperature even down to the lowest measured value of $T$. Finally, the kink in $\Imone$ and the inflection point in $\sone$ occur at $\Tstar\approx0.4~\mathrm{K}$. These features all indicate a stronger relevance of PD, such that it is present at $20~\mathrm{mK}$ and becomes dominant at a lower temperature: $\Tstar$ is decreased. 
\setcounter{myc}{7}

\setcounter{myc}{8}
\begin{figure}
	\includegraphics[width=\columnwidth]{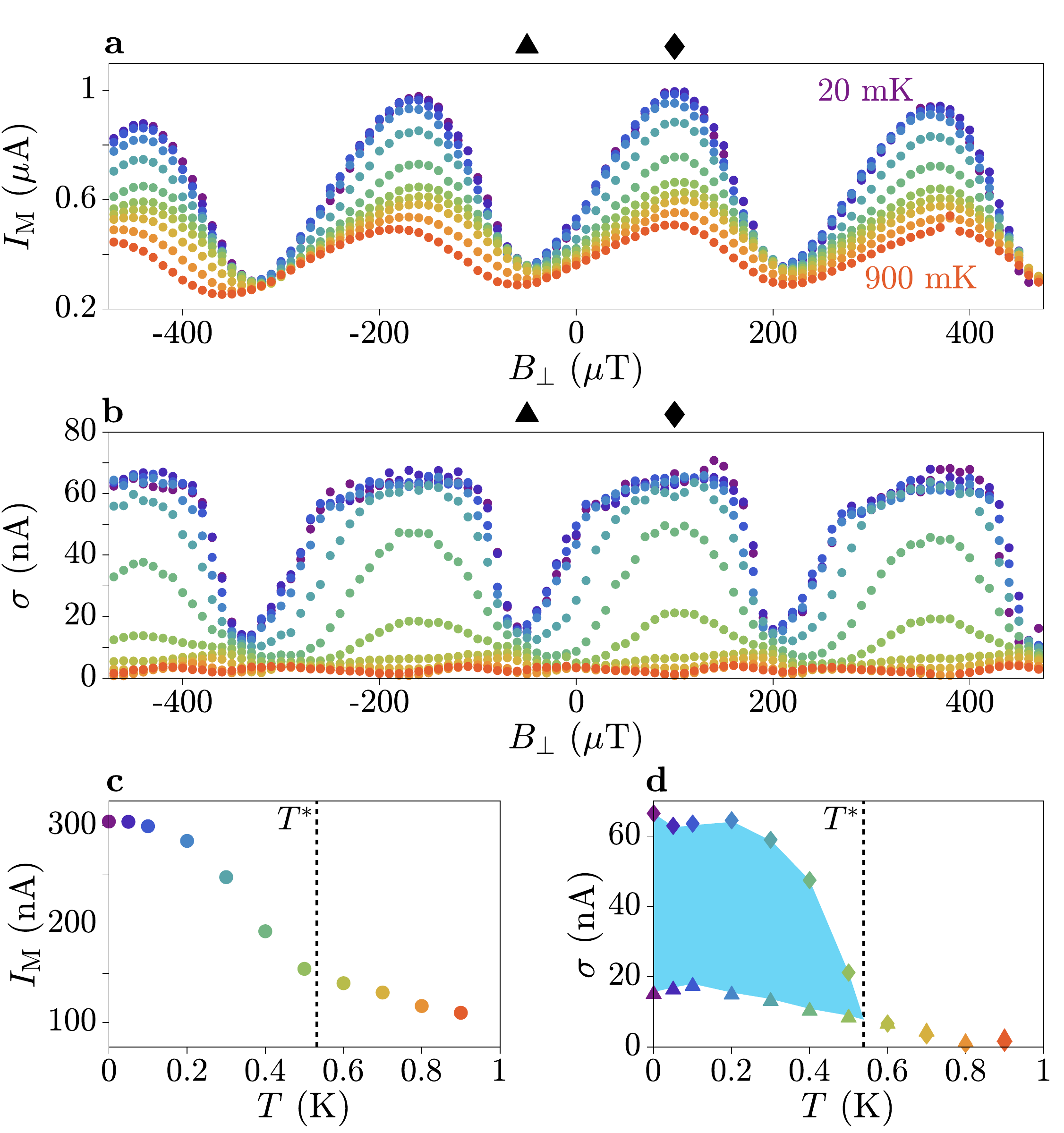}
	\caption{(a) Mean switching current $\ImS$ for the SQUID configuration $\Vgone=-300~\mathrm{mV}$ and $\Vgtwo=-140~\mathrm{mV}$, as a function of $\Bperp$. Field traces are taken at temperatures ranging from $20~\mathrm{mK}$ up to $900~\mathrm{mK}$. The maximum (diamond) and minimum (triangle) of the $\ImS$ oscillations are marked. (b) Standard deviation $\sS$ in this SQUID configuration, for temperatures $20~\mathrm{mK}$ to $900~\mathrm{mK}$. (c) Oscillation amplitude of $\ImS$, $\Delta\ImS/2$ as a function of temperature. A kink in $\Delta\ImS/2$ occurs at the transition temperature of $T^{*}\approx0.52~\mathrm{K}$. (d) Standard deviation $\sS$ at values of $\Bperp$ corresponding to the $\ImS$ maximum (diamonds) and minimum (triangles) respectively. The large difference at $T<T^{*}$ is indicated.}
	\label{Sfig5}
\end{figure}

The regime of the SQUID is dominated by JJ1, the large $\Ic$ component, so we can change the SQUID behavior by varying $\Vgone$. For $\Vgone=-300~\mathrm{mV}$, the SQUID undergoes direct transitions between MQT and PD depending on $\Bperp$ [Fig.~4(b) in the Main Text]. The full dataset is shown in Fig.~\ref{Sfig5} for $20~\mathrm{mK}$ to $900~\mathrm{mK}$: the mean switching current $\ImS$ in (a) and the standard deviation in (b). 

The oscillation amplitude $\Delta\ImS/2$ is shown in Fig.~\ref{Sfig5}(c). The enhancement in switching current at low temperatures is again observed, where quantum tunneling is dominant. We also observe the characteristic kink in $\Delta\ImS/2$, in this case at $T\approx0.52~\mathrm{K}$ concomitant with the lower transition temperature to the phase diffusive regime.

At $T=20~\mathrm{mK}$, we observe a large variation in the standard deviation $\sS$ depending on the field $\Bperp$. At the maximum of $\ImS$ (diamond), $\sS$ is large at low temperature. This is consistent with quantum tunneling as the dominant mechanism of phase escape. Instead, $\sS=20~\mathrm{nA}$ at the minimum (triangle), indicating that phase diffusive effects are strong. The traces in $\sS$ at these field values are shown in Fig.~\ref{Sfig5}(d) by their respective markers. The large difference in $\sS$ is evident at low temperatures, as indicated by the blue shading, where the external magnetic field determines the extent of phase diffusion in the SQUID. On increasing $T$ towards the transition temperature, the difference in $\sS$ reduces until the SQUID is fully phase-diffusive at all values of $\Bperp$. 

\setcounter{myc}{9}
\begin{figure}
	\includegraphics[width=\columnwidth]{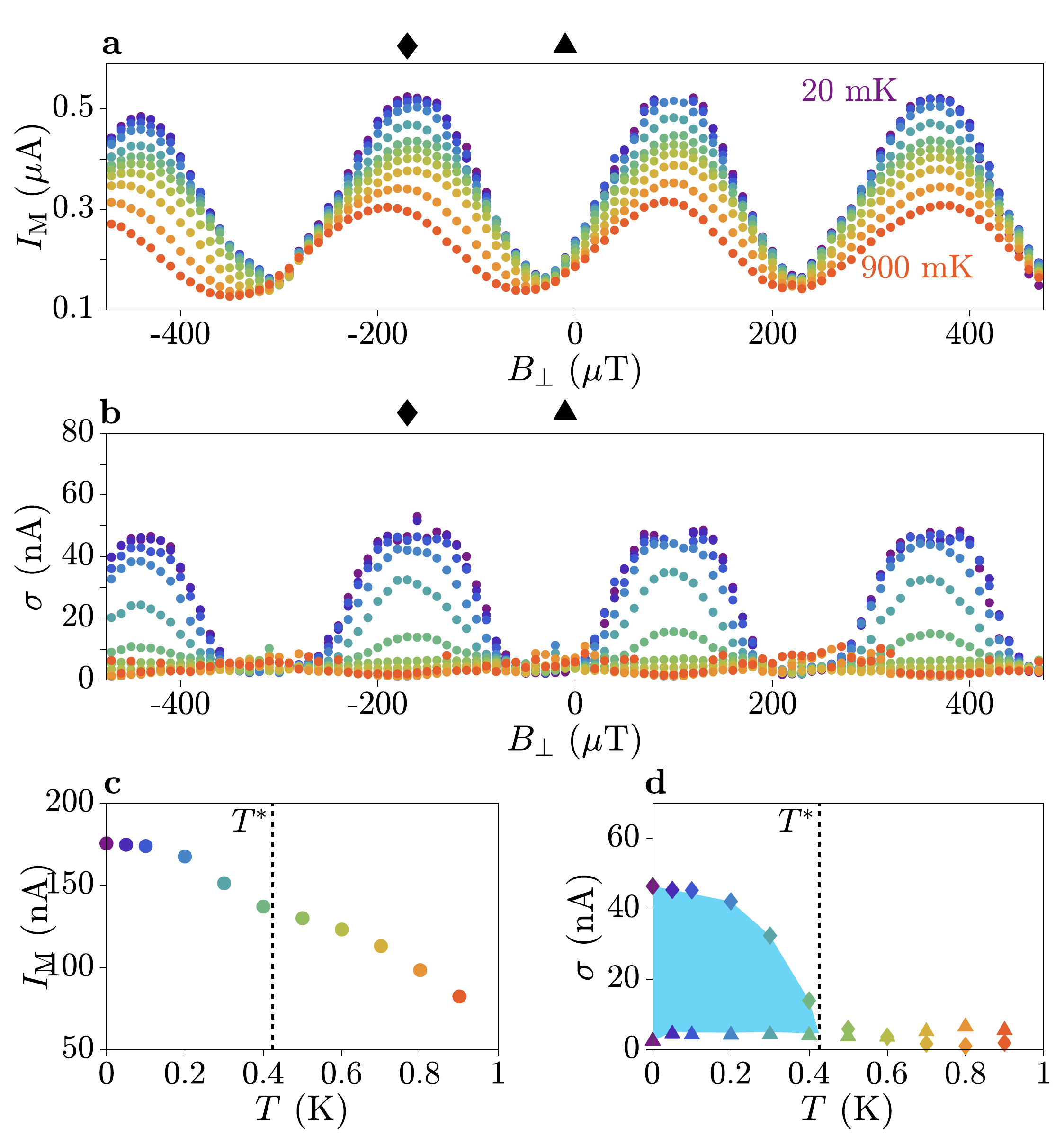}
	\caption{(a) Mean switching current $\ImS$ for the SQUID configuration $\Vgone=-350~\mathrm{mV}$ and $\Vgtwo=-140~\mathrm{mV}$, as a function of $\Bperp$. Field traces are taken at temperatures ranging from $20~\mathrm{mK}$ up to $900~\mathrm{mK}$. The maximum (diamond) and minimum (triangle) of the $\ImS$ oscillations are marked. (b) Standard deviation $\sS$ in this SQUID configuration, for temperatures $20~\mathrm{mK}$ to $900~\mathrm{mK}$. (c) Oscillation amplitude of $\ImS$, $\Delta\ImS/2$ as a function of temperature. A kink in $\Delta\ImS/2$ occurs at the transition temperature of $T^{*}\approx0.42~\mathrm{K}$. (d) Standard deviation $\sS$ at values of $\Bperp$ corresponding to the $\ImS$ maximum (diamonds) and minimum (triangles) respectively. The large difference at $T<T^{*}$ is indicated.}
	\label{Sfig6}
\end{figure}

\setcounter{myc}{10}
\begin{figure}
	\includegraphics[width=\columnwidth]{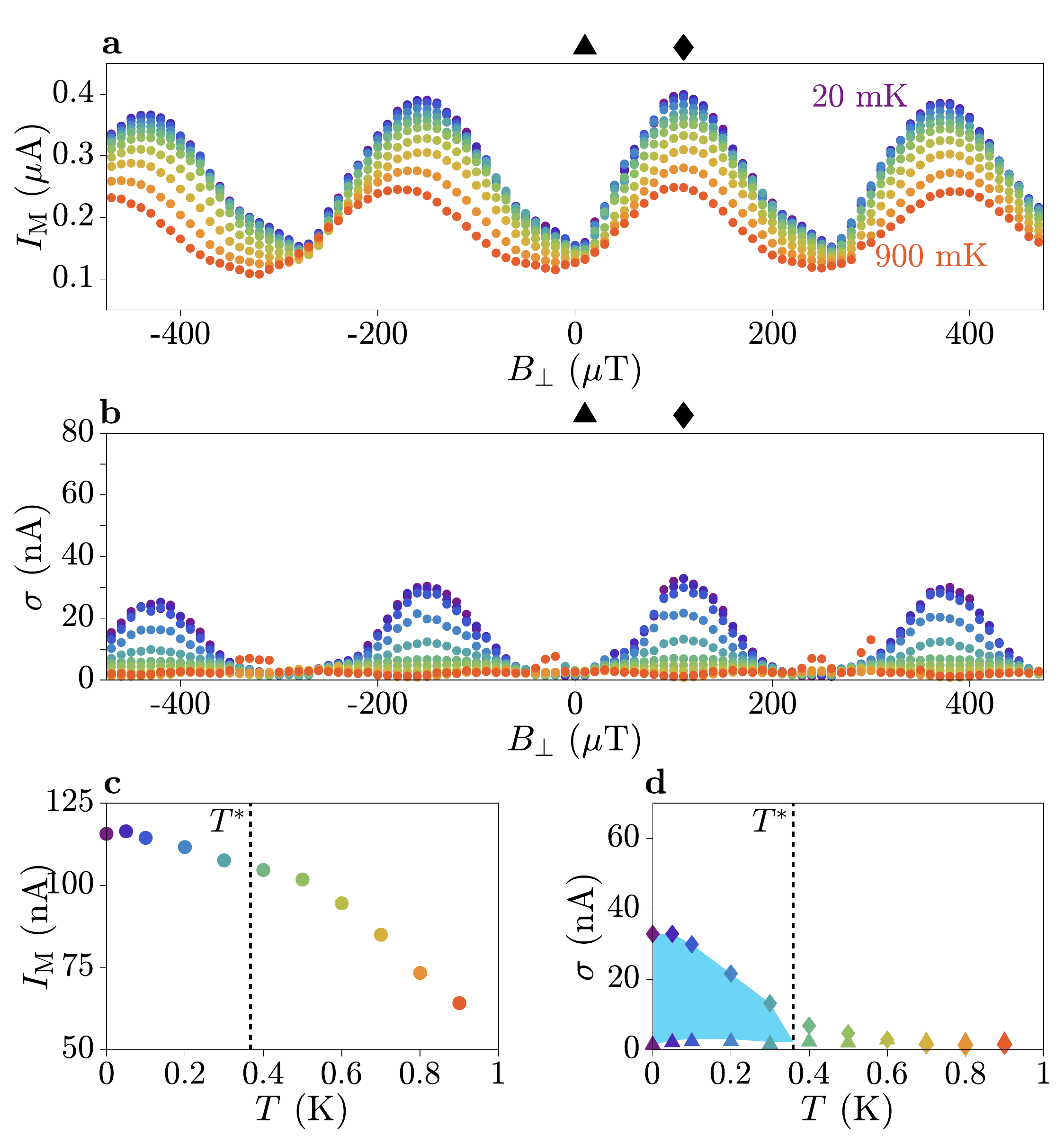}
	\caption{(a) Mean switching current $\ImS$ for the SQUID configuration $\Vgone=-375~\mathrm{mV}$ and $\Vgtwo=-140~\mathrm{mV}$, as a function of $\Bperp$. Field traces are taken at temperatures ranging from $20~\mathrm{mK}$ up to $900~\mathrm{mK}$. The maximum (diamond) and minimum (triangle) of the $\ImS$ oscillations are marked. (b) Standard deviation $\sS$ in this SQUID configuration, for temperatures $20~\mathrm{mK}$ to $900~\mathrm{mK}$. (c) Oscillation amplitude of $\ImS$, $\Delta\ImS/2$ as a function of temperature. A kink in $\Delta\ImS/2$ occurs at the transition temperature of $T^{*}\approx0.35~\mathrm{K}$. (d) Standard deviation $\sS$ at values of $\Bperp$ corresponding to the $\ImS$ maximum (diamonds) and minimum (triangles) respectively. The large difference at $T<T^{*}$ is indicated.}
	\label{Sfig7}
\end{figure}

Figures~\ref{Sfig6} and \ref{Sfig7} show the datasets for $\Vgone=-350~\mathrm{mV}$ and $\Vgone=-375~\mathrm{mV}$, respectively. While still asymmetric for $\Vgone=-350~\mathrm{mV}$, the critical current of JJ1 is no longer much larger than that of JJ2. The much lower critical current of JJ1 means that quantum tunneling is only dominant close to the maximum of the SQUID oscillations. This is highlighted in Fig.~\ref{Sfig6}(b). At the minima, the SQUID is fully phase diffusive with $\sS<5~\mathrm{nA}$. The oscillation amplitude $\Delta\ImS/2$ in Fig.~\ref{Sfig6}(c) shows a kink at $T\approx0.42~\mathrm{K}$, which is consistent with the average transition temperature between MQT and PD regimes. Figure~\ref{Sfig6}(d) shows $\sS$ at flux values corresponding to the maximum (diamonds) and minimum (triangles) of the $\ImS$ oscillations. The divergence for $T<T^{*}$ is evident, as in Fig.~\ref{Sfig5}(d). We note that at some values of $\Bperp$ in the low-temperature curve the switching current went above the signal amplitude, artificially truncating the SPD and rendering its standard deviation unphysical: these points have been removed from the trace.

On further decrease in $\Vgone$ to $-375~\mathrm{mV}$, the SQUID is almost symmetric. In this case, a magnetic-field dependence is still observable in the standard deviation [see Fig.~\ref{Sfig7}(b)] but the SQUID is phase diffusive at $T=20~\mathrm{mK}$ for all values of $\Bperp$. The corresponding oscillation amplitude, while no longer representative of the switching current of JJ2, again shows the kink in $\Delta\ImS/2$ at the low transition temperature of $T\approx0.35~\mathrm{K}$. Figure~\ref{Sfig7}(d) shows $\sS$ at the maxima (diamonds) and minima (triangles) of $\ImS$, and while some divergence emerges for $T<T^{*}$, phase diffusion is dominant for all values of $\Bperp$ .

\end{document}